%% file: conference_101719.tex
\documentclass[conference]{IEEEtran}

\usepackage[english]{babel}
\usepackage{cite}
\usepackage{amsmath,amssymb,amsfonts}
\usepackage{graphicx}
\usepackage{textcomp}
\usepackage{xcolor}
\usepackage{booktabs}
\usepackage{enumitem}
\usepackage{listings}
\usepackage{multicol}
\usepackage{multirow}
\usepackage{array}
\usepackage{url}
\usepackage{tcolorbox}
\usepackage{stfloats}
\usepackage[commandnameprefix=always]{changes}

\definecolor{BrickRed}{RGB}{203,65,84}
\definecolor{ForestGreen}{RGB}{34,139,34}
\tcbuselibrary{listings, skins, breakable}
\providecommand{\citep}[1]{\cite{#1}}
\providecommand{\citet}[1]{\cite{#1}}

\def\BibTeX{{\rm B\kern-.05em{\sc i\kern-.025em b}\kern-.08em
    T\kern-.1667em\lower.7ex\hbox{E}\kern-.125emX}}

\begin{document}

\title{UModel: An Agent-Ready Observability Data Modeling Method at Scale}


\author{
Changhua Pei\IEEEauthorrefmark{1}\IEEEauthorrefmark{3},
Zheyuan Li\IEEEauthorrefmark{1}\IEEEauthorrefmark{2},
Zexin Wang\IEEEauthorrefmark{1}\thanks{Corresponding author: Zexin Wang (wangzexin@cnic.cn).},
Hang Cui\IEEEauthorrefmark{1}\IEEEauthorrefmark{2},
Xiaohui Nie\IEEEauthorrefmark{1}\IEEEauthorrefmark{3},
Qi Zhou\IEEEauthorrefmark{4},
Fang Situ\IEEEauthorrefmark{4},
Cheng Zhang\IEEEauthorrefmark{4},\\
Xin Zhang\IEEEauthorrefmark{4},
Xidao Wen\IEEEauthorrefmark{4},
Gaogang Xie\IEEEauthorrefmark{1}\IEEEauthorrefmark{2},
Jingjing Li\IEEEauthorrefmark{1},
Dan Pei\IEEEauthorrefmark{5}
\\[0.5em]
\IEEEauthorrefmark{1}CNIC, CAS
\IEEEauthorrefmark{2}UCAS 
\IEEEauthorrefmark{3}Hangzhou Institute for Advanced Study, UCAS\\
\IEEEauthorrefmark{4}Alibaba 
\IEEEauthorrefmark{5}Tsinghua University
}

\maketitle

\begin{abstract}
When networked system failures occur, automatically performing Root Cause Analysis (RCA) using observability data is critical for ensuring networked system reliability. Recently, LLM-based agents have shown promise for automating this diagnosis process through advanced reasoning and autonomous exploration. However, existing observability frameworks remain archaic, characterized by fragmented data silos, incompatible schemas, and insufficient semantic metadata, preventing agents from establishing the complex relationships required for effective RCA. To address these challenges, we present \textit{UModel}, a unified ontological framework that shifts observability from data-centric to object-centric modeling. UModel constructs a virtual ontological layer where heterogeneous telemetry, entities, and expert knowledge are standardized as objects and interconnected via semantic graphs. In addition, we introduce U-SPL, a pipeline-based query interface that enables agents to autonomously explore system topologies and correlate multimodal data. By re-modeling the ``AIOps 2025 Challenge'' dataset using UModel, the precision of root cause localization improved by 8\%, demonstrating that enhanced data organization can significantly increase the accuracy of downstream tasks. UModel provides a scalable modeling framework that, in its deployment at Alibaba Cloud for more than one year, has served tens of thousands of users, sustained millions of operations per second, and delivered sub-second query latency.
\end{abstract}


\input{body}

\bibliographystyle{IEEEtran}
\bibliography{reference}

\appendix
\label{sec:appendix}

\input{appendix}

\end{document}

%% file: body.tex
\section{Introduction}
\label{sec:intro}

\input{introduction_v2}

\section{Preliminaries}
\label{sec:preliminaries}

\input{preliminary}

\section{Agent-Ready Data Model}
\label{sec:agent_ready_data_model}
\input{agent_ready_data_model_v2}

\section{UModel Architecture and Design}
\label{sec:design}

\input{method_v2}



\section{Production Deployment and Lessons}

\input{deploymentandlessons}
\label{deploymentandlessons}

\section{Evaluation}
\label{sec:realworld}
\input{realworlddeployment}

\section{Related Work}
\label{sec:Related Work}

\input{RelatedWork}

\section{Conclusion}
\label{sec:conclusion}

\input{conclusion}



%% file: introduction_v2.tex
The rapid proliferation of Large Language Models (LLMs) and multi-agent systems has fundamentally transformed the landscape of software and networked system engineering. Today, the cost of constructing complex systems from scratch has plummeted. Tools such as Replit~\cite{replit_2024} and Google's Antigravity Tool~\cite{google_antigravity_2025} allow developers to build sophisticated distributed applications with unprecedented speed. However, while the barrier to \textit{creation} has lowered, the complexity of ensuring networked system stability and reliability has grown exponentially. Consequently, the advancement of operational capabilities, specifically in Site Reliability Engineering (SRE), has failed to keep pace with the velocity of system construction.


In recent years, Artificial Intelligence for IT Operations (AIOps)~\cite{gartner_aiops_2016,gartner_aiops_maturity_2025,diaz_de_arcaya_2023_aiops_mlops_survey,zhang_2024_aiops_llm_survey} has emerged as a promising solution to bridge this gap. By applying machine learning to observability data, AIOps aims to automate the operational lifecycle from anomaly detection~\cite{xu2018unsupervised,drain2017,xie2023point,zhou2025kanad} to Root Cause Analysis~\cite{li2022actionable,soldani2022anomaly,zhang2025failure,yu2026survey} (RCA). Currently, anomaly detection within single modalities, such as metrics, logs, and traces, has achieved significant maturity, enabling accurate identification of anomalies. However, RCA, which is essential for pinpointing the fundamental origins of faults and guiding system restoration or optimization, remains a significant bottleneck. Existing RCA methods often lack the precision required for large-scale production deployment.

Recent research has explored multimodal fusion techniques~\cite{yu2023nezha,zhang2025failure}. However, these methods have yielded only marginal improvements without achieving a fundamental breakthrough. The core limitation lies in their reliance on supervised learning paradigms, which require the model to learn correlations between system failures and root causes based on historical data. This approach relies on ``memorization'', which proves brittle for ensuring networked system reliability. Studies indicate that over 40\% of faults in real-world industrial scenarios are zero-shot failures that the model has never encountered~\cite{li2022actionable}. Consequently, the performance of ``memory-based'' models degrades sharply in these unseen scenarios.



LLM-based agents have introduced a new paradigm for autonomously accomplishing AIOps tasks \cite{wang2024rcagent,pei2025flow}. By leveraging their extensive general knowledge, these agents are able to perform automated reasoning and execute actions, enabling unsupervised operation in complex environments. However, existing observability data models are not yet agent-ready and primarily face two key challenges:

\begin{enumerate}
    \item \textbf{Deficiency of Critical Information and Relationships:} 
    Logs, metrics, traces, and events are scattered across disparate storage systems with incompatible schemas. This fragmentation results in a severe lack of metadata: systems merely store raw data without recording its semantic utility, intended usage, or processing methods. Furthermore, current models fail to establish the complex multi-dimensional relationships required for effective RCA. 

    \item \textbf{Lack of Support for Autonomous Agent Learning:} 
    Current approaches rely heavily on human operators to filter and pre-process data based on their subjective understanding of what is ``useful'' before feeding it to the model. However, manual pre-processing often results in significant information loss and renders the data pipeline unscalable. 
    The correct paradigm should be that the agent can actively explore the system, autonomously retrieving relevant topology and context on demand, thereby ensuring both real-time accuracy and adaptability.
\end{enumerate}

Addressing these challenges is technically demanding. It requires unifying heterogeneous telemetry from incompatible systems such as Prometheus \cite{prometheus} and Elasticsearch \cite{elasticsearch}. This process involves rigorous schema alignment rather than simple data transport. To transform ``dead'' values into agent-readable insights, the system must build a comprehensive ontology that makes data self-describing, replacing implicit human intuition. This complexity is amplified by the need to map implicit dependencies into a real-time knowledge graph to track fault propagation. Rigid legacy interfaces must be superseded by a unified protocol, enabling agents to perform iterative, context-aware exploration at scale.



To address these challenges, we introduce a unified object-centric model, referred to as \textbf{UModel}, which standardizes all operational resource types—including instances, data, and knowledge—into \texttt{objective sets}. Under this abstraction, every element in UModel, whether a microservice instance, a telemetry signal (e.g., metrics, logs, or traces), or a standard operating procedure (e.g., a handbook), is represented as an object. Objects are interconnected via \texttt{links} (e.g., \textit{entity--data links}), forming a semantic graph that bridges the gap between raw observational data and operational logic. This standardization enables agents to more effectively capture intrinsic relationships among entities, thereby improving RCA results. Furthermore, we design a \textbf{U}nified \textbf{S}earch \textbf{P}rocessing \textbf{L}anguage (\textbf{U-SPL}), which encapsulates heterogeneous entities and underlying resources within a unified, pipeline-oriented query interface. The object-oriented design of U-SPL makes it inherently more agent-ready. By providing a unified syntax, U-SPL significantly reduces the learning overhead for agents. Its minimalist query formulation enables agents to execute complex queries—such as those involving joins across multiple heterogeneous systems—with low operational cost and a reduced likelihood of errors.



We have deployed UModel in the production environment of Alibaba Cloud for more than one year. UModel has served tens of thousands of users for an extended period. The system has achieved massive scale, unifying thousands of standardized models across numerous heterogeneous subsystems (e.g., K8s, Network) and covering critical domains such as Container Services, APM, and Cloud Infrastructure Monitoring. Despite this complexity, the system maintains industrial-grade performance, handling millions of operations per second with second-level query response times for massive data retrieval. To rigorously evaluate its efficacy in RCA, we modeled the ``2025 AIOps Challenge''~\cite{sun2025aiopsarena} dataset using UModel. Experiments demonstrate that equipped with our UModel, the naive agent-based approach improves RCA accuracy by 8\%. 

The contributions of this paper are as follows:
\begin{itemize}
    \item We propose \textbf{UModel}, a unified ontological framework that standardizes the representation of heterogeneous operational data, entities, and knowledge, enabling zero-shot transferability for AIOps agents.
    \item We introduce \textbf{U-SPL}, a unified interface that allows LLM-based agents to autonomously explore system topologies and correlate multimodal data without rigid, hand-tuned prompts.
    \item We present empirical evidence from large-scale industrial deployment at Alibaba and benchmark performance on the ``2025 AIOps Challenge'' dataset, demonstrating significant enhancement in root cause localization.
\end{itemize}


%% file: preliminary.tex
\subsection{Search Processing Language}

\textbf{SPL Syntax and Composition.} At its core, an AIOps task consists of retrieving and analyzing operational data. To automate this process, we adopt the Search Processing Language (SPL), a pipeline-oriented query language conceptually inspired by Unix pipes \cite{kernighan1979unix}. SPL enables progressive data processing, where intermediate results are incrementally transformed and filtered through a sequence of composable operators, supporting scalable and modular analysis of heterogeneous telemetry.

An SPL query is composed of a \textbf{Source} followed by a series of \textbf{Pipe Commands}, separated by the vertical bar symbol (\texttt{|}).
\begin{equation}
\small
    \texttt{Source} \xrightarrow{} \texttt{|} \ \texttt{Command}_1 \xrightarrow{} \texttt{|} \ \texttt{Command}_2 \xrightarrow{} \dots
\end{equation}
The output of the previous command becomes the input for the next, allowing for complex data transformations to be built from simple, modular blocks.

\textbf{A ``Data Assembly Line'' Example.} To illustrate SPL, consider a scenario where an operator needs to find the error rate of a specific microservice. We can visualize this as an assembly line processing raw materials (logs):

\begin{enumerate}
    \item \textbf{Source (Retrieval):} First, we grab all raw logs from the storage.
    \item \textbf{Filter (Screening):} We filter out everything except errors from the ``checkout'' service.
    \item \textbf{Stats (Aggregation):} We count these errors grouped by time to see the trend.
    \item \textbf{Action (Anomaly Detection):} Use anomaly detection tool to detect the service's anomaly.
\end{enumerate}

The corresponding SPL query is written as follows:

{\small
\begin{verbatim}
* | where service = "checkout" and level = "ERROR" 
| stats count() by bin(timestamp, 1m)
| anomaly-detection 
\end{verbatim}
}

In this example, \texttt{*} represents the initial raw data stream. The first pipe (\texttt{|}) passes this stream to the \texttt{where} command, which acts as a sieve, discarding irrelevant logs. The surviving data is passed through the second pipe to the \texttt{stats} command, which compresses the data into a statistical summary. This pipeline approach allows users (and agents) to construct sophisticated analytical logic step-by-step.

%% file: agent_ready_data_model_v2.tex

Before delving into the deficiencies of existing operational data and introducing our UModel, we first establish the definition and organization criteria for ``agent-ready'' operational data. To transform raw operational data into knowledge that an AI agent can reason about and act upon, the data architecture must adhere to the following four core principles:
\begin{itemize}
    \item \texttt{Rule 1: Semantically Rich.} (Section~\ref{sec:gap_semantics})  
    
    \item \texttt{Rule 2: Contextualized.} (Section \ref{sec:gap_relationships}) 
    
    \item \texttt{Rule 3: Actionable.} (Section \ref{sec:gap_tools})
    
    \item \texttt{Rule 4: Structured \& Standardized.} (Section~ \ref{sec:gap_standards}) 
\end{itemize}

In the following, we analyze how traditional observability stacks fail each criterion and how UModel addresses them.


\subsection{Gap in Semantics: From Dead Tokens to Self-Describing Objects}
\label{sec:gap_semantics} 

In traditional observability, data exists as ``dead'' strings and numbers optimized for storage rather than reasoning. For example, a metric named \texttt{node\_cpu\_seconds\_total} carries implicit meaning for a human operator, who relies on experience to understand its unit (seconds), its type (counter), and the necessity of applying a rate function to derive utility. However, to an LLM-based agent, this identifier is merely a token string devoid of context.

This lack of self-describing metadata creates a severe barrier for autonomous reasoning. Without explicit definitions regarding units and data types, agents struggle to construct valid queries (such as PromQL), often leading to execution errors or logical hallucinations. To address this, an agent-ready model must implement a robust \textbf{Semantic Layer}. Data must be self-describing, carrying enriched metadata that explicitly informs the agent ``what this data is'' and ``how it should be queried'', thereby replacing implicit human intuition with explicit machine-readable context.

\subsection{Gap in Relationships: From Isolated Events to Causal Graphs}
\label{sec:gap_relationships} 

Traditional models predominantly store time-series data and discrete events, capturing ``what happened'' but failing to capture the structural ``why''. Human operators bridge this gap by reconstructing dependency graphs from architecture diagrams. Agents, lacking this external knowledge, view logs and metrics as flat, unrelated lists. Without an explicit topological structure, an agent cannot logically deduce that a failure in service B is the direct causal result of an outage in service A; it is limited to observing statistical correlations.

To enable true reasoning, the data model must be \textbf{Graph-Based}, evolving into an observable knowledge graph. In this paradigm, entities are nodes and dependencies are explicit edges (e.g., \texttt{Node A --[calls]--> Node B}). This structure allows the agent to traverse the system topology, effectively ``following the vine to find the melon'', rather than guessing at relationships. By embedding the topology directly into the data model, we enable the agent to perform multi-hop reasoning and precise root cause isolation.




\subsection{Gap in Tools: Enabling Action and Data Pre-processing}
\label{sec:gap_tools} 

Most current AIOps systems function as read-only dashboards, creating a disconnect between ``seeing'' and ``doing''. This gap in tools is twofold: it hinders both the output (action) and the input (perception) of the agent. 

First, regarding action, seeing a ``DiskFull'' alert without a bound execution path (e.g., a \texttt{cleanup.sh} script) prevents the agent from closing the operational loop. Second, and equally critical, is the lack of data pre-processing tools. Feeding raw, massive logs directly to an agent creates a low signal-to-noise ratio that overflows the context window.

Therefore, an agent-ready model must be \textbf{Tool-Enabled}. It must provide a suite of tools that serve two purposes:
\begin{enumerate}
    \item \textbf{Action Execution:} Binding entities to specific remediation tools to enable autonomous recovery.
    \item \textbf{Data Pre-processing:} Providing statistical and processing tools (e.g., \texttt{filter}, \texttt{aggregate}, \texttt{rank}) that allow the agent to distill gigabytes of raw telemetry into concise insights \textit{before} ingesting them into the LLM context. This ``tool-use before reading'' strategy is essential for handling scale.
\end{enumerate}

\subsection{Gap in Heterogeneity: Bridging Silos}
\label{sec:gap_standards} 
Operational data is notoriously fragmented across heterogeneous systems, with metrics, logs, traces, and events stored in isolated silos such as Prometheus, ELK, and Jaeger. This fragmentation forces the agent to assume the role of an ETL (Extract, Transform, Load) engineer. To form a complete system view, an agent must query multiple disparate APIs, handle conflicting protocols, and manually reconcile inconsistent identifiers—such as mapping a \texttt{service\_id} in logs to a \texttt{serviceName} in traces.

An agent-ready model must function as a \textbf{Unified Object Store}. It requires a consistent schema that standardizes heterogeneous entities, regardless of their underlying storage implementation. Whether the underlying log storage is Elasticsearch or Loki, or the metric store is Prometheus or VictoriaMetrics, the agent interacts with a unified ``entity'' abstraction. This standardization eliminates the need for the agent to learn specific dialects for every tool and significantly reduces the probability of hallucination during context stitching, allowing the agent to focus reasoning power on diagnosis rather than data integration.

%% file: method_v2.tex

\subsection{Overall Design}

Based on the definition of an agent-ready data model in Section \ref{sec:agent_ready_data_model}, the design of UModel is anchored in two core architectural pillars:

\subsubsection{Pillar 1: Object-Centric Semantic Organization}
UModel abandons the data-centric view in favor of an \textbf{Object-Centric} paradigm. All operational assets, whether they are physical resources (Hosts, Pods), telemetry data (Logs, Metrics), or knowledge assets (Runbooks), are standardized as \textbf{Entities}.
\begin{itemize}
    \item \textbf{Unified Graph Structure:} These entities do not exist in isolation. They are interconnected via \textbf{Links} that explicitly represent topological, logical, and data dependencies.
    \item \textbf{Semantic \& Knowledge Attachment:} Through a standardized schema design, we attach high-level semantics (natural language descriptions), expert knowledge (diagnosis rules), and tools (remediation scripts) directly to these entities.
\end{itemize}
This transforms the system from a ``database of numbers'' into an environment that the agent can reason about logically.

\subsubsection{Pillar 2: Unified Interface and Protocol}
To prevent the agent from needing to learn the dialects of heterogeneous underlying systems (e.g., Prometheus vs. Elasticsearch), UModel provides a \textbf{Unified Interaction Protocol}.
\begin{itemize}
    \item \textbf{Standardized Abstraction:} We decouple the logical model from physical storage. The agent interacts with a consistent ``entity'' abstraction, while UModel handles the translation to underlying storage engines.
    \item \textbf{One-Stop Exploration (SPL):} A unified SPL provides the unified interface for all query types. This guarantees that agents can easily explore the system, discovering entities, and traversing relationships.
\end{itemize}



\begin{figure}[t]
  \centering
  \includegraphics[width=\linewidth]{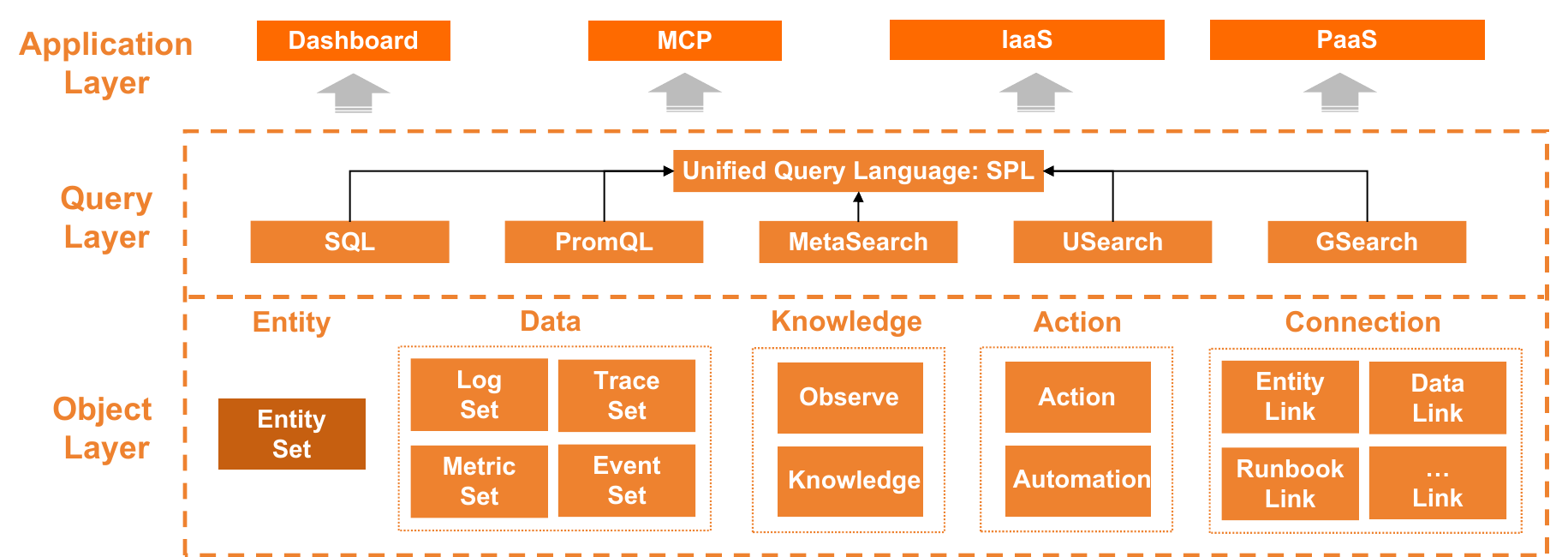}
  \caption{Overview of UModel.}
  \label{fig:model}
\end{figure}

Guided by the overall design described above, Figure~\ref{fig:model} presents an overview of the UModel architecture, which is organized into three conceptual layers: the object layer, the query layer, and the application layer. At the application layer, we implement a dashboard as well as IaaS and PaaS interfaces, and we further deploy an Model Context Protocol (MCP) server. In the following, we focus on a detailed description of three layers.


\subsection{Object Layer: Object-Centric Modeling}

\begin{figure}[t]
  \centering
  \includegraphics[width=\linewidth]{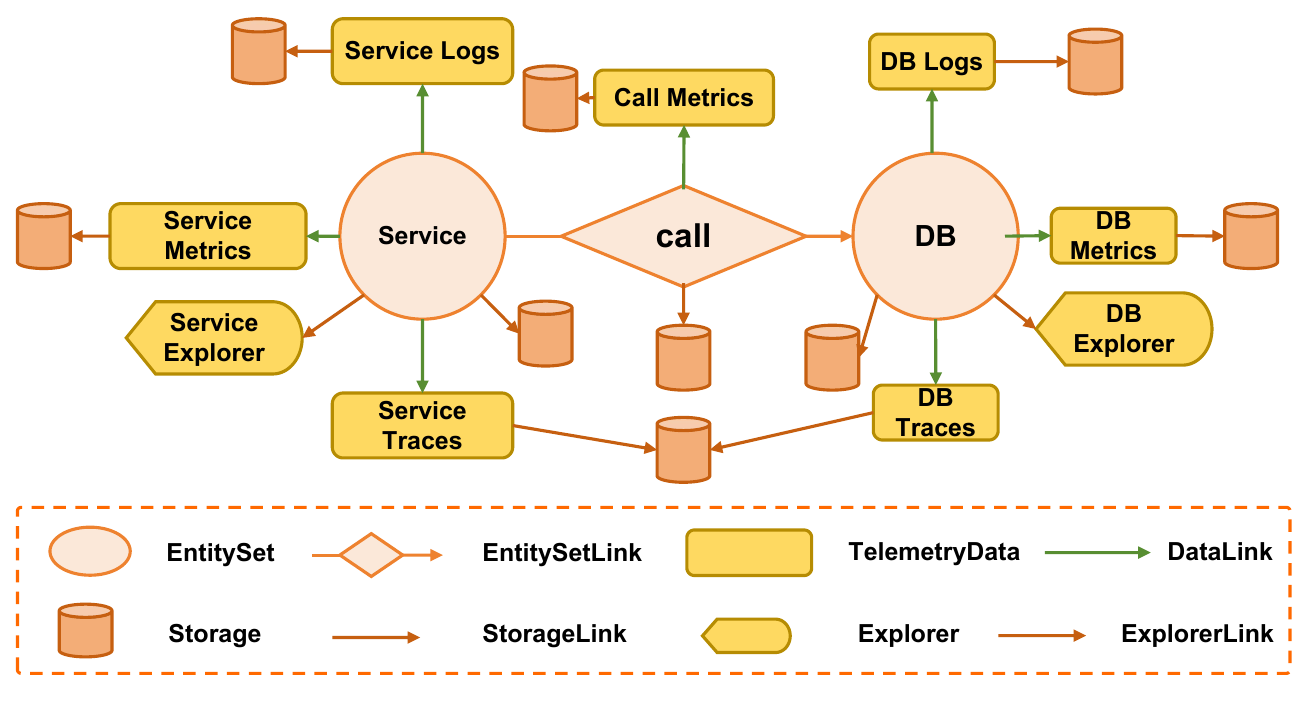}
  \caption{An example of UModel.}
  \label{fig:umodel-sample}
\end{figure}

UModel fundamentally transforms the traditional observability landscape by shifting from a data-centric model to an object-centric graph paradigm. As illustrated in Figure~\ref{fig:umodel-sample}, the system constructs a digital twin of the operational environment where resources are abstracted as nodes (EntitySet) and their interdependencies as edges (EntitySetLink). In this architecture, a \textit{Service} entity is not merely a label in a log line but a distinct node in a relational network, explicitly connected to a \textit{DB} entity via a \textit{call} relationship. Figure~\ref{fig:umodel-sample} further demonstrates that heterogeneous telemetry data, spanning \textit{ServiceMetrics}, \textit{ServiceLogs}, and \textit{ServiceTraces}, are no longer fragmented silos. Instead, they are unified through DataLinks that bind specific datasets directly to their corresponding entities. This structure allows an AI agent to navigate the system topology seamlessly, moving from a service to its dependencies without manual context stitching.

To clarify the logical relationships between the topology and the underlying data, we have summarized the core concepts of UModel and their respective significance in Table \ref{tab:umodel_concepts}. The UModel architecture is built upon six pillars: EntitySet and EntitySetLink form the skeleton of the system, defining the business objects and their topological dependencies. DataSet provides standardized schemas for heterogeneous telemetry (Metrics, Logs, and Traces), while Storage handles the physical persistence layer. Finally, DataLink and StorageLink serve as critical connectors. The former logically binds observational data to specific entities, while the latter physically routes these datasets to their appropriate storage engines. This decoupling ensures that the modeling graph remains semantically rich while agnostic to the underlying storage implementation.

\begin{table}[t]
    \centering
    \footnotesize
    \caption{Core concepts and attributes of UModel.}
    \label{tab:umodel_concepts}
    \renewcommand{\arraystretch}{1.5}
    \resizebox{\linewidth}{!}{
    \begin{tabular}{p{0.18\linewidth} p{0.4\linewidth} p{0.4\linewidth}}
        \toprule
        \textbf{Concept} & \textbf{Function} & \textbf{Key Schema} \\
        \midrule

        \textbf{EntitySet} &
        Defines the structure and properties of a business object. Acts as the ``template''. &
        \texttt{metadata.name}, \newline
        \texttt{spec.fields}, \newline
        \texttt{spec.primary\_key\_fields} \\
        \midrule

    \textbf{DataSet} & Defines the schema for telemetry data (\textbf{Metrics, Logs, Traces, and Events}). & Metric: metrics, labels \newline Log: fields, message\_field \newline Trace: trace\_id\_field \newline Event: event\_type, message \\
    \midrule
    


        \textbf{EntitySetLink} &
        Defines topological relationships between entities. &
        \texttt{left\_entity\_set}, \newline
        \texttt{right\_entity\_set}, \newline
        \texttt{link\_type}  \\
        \hline
        \textbf{DataLink} &
        Logically binds a DataSet to an EntitySet via field mapping. &
        \texttt{src} (Entity), \newline
        \texttt{dest} (DataSet)\\

        \bottomrule
    \end{tabular}
    }
\end{table}

To implement this graph structure in Figure~\ref{fig:umodel-sample} rigorously, UModel employs comprehensive schema configurations as detailed in Table~\ref{tab:umodel_concepts}. The \textit{EntitySet Schema} establishes the identity of graph nodes by defining `spec.fields' and enforcing uniqueness through `spec.primary\_key\_fields'. Concurrently, the topology is formalized using the \textit{EntitySetLink Schema}, which captures the business semantics of edges. By utilizing the `spec.link\_type' field to specify relationships such as \textit{calls}, \textit{contains}, or \textit{hosts}, and defining precise field mappings between the `left\_entity\_set' and `right\_entity\_set', UModel converts implicit system knowledge into an explicit, traversable knowledge graph.

For observational data, as shown in Table~\ref{tab:unified_set_config}, UModel represents telemetry not as unstructured numbers or strings, but as structured objects defined through \textit{MetricSet}, \textit{LogSet}, and \textit{TraceSet} configurations. For example, the \textit{TraceSet Schema} standardizes distributed tracing by requiring fields such as \textit{trace\_id\_field} and \textit{parent\_span\_id\_field}. By binding these self-describing datasets to the entity graph, UModel enables an agent exploring the \textit{Service} node in Figure~\ref{fig:umodel-sample} to obtain immediate and semantically consistent access to all relevant operational data, thereby bridging the gap between underlying telemetry and agent reasoning.

\subsection{Query Layer: U-SPL}




\begin{figure*}[htbp]
\centering
\centering  \includegraphics[width=0.9\linewidth]{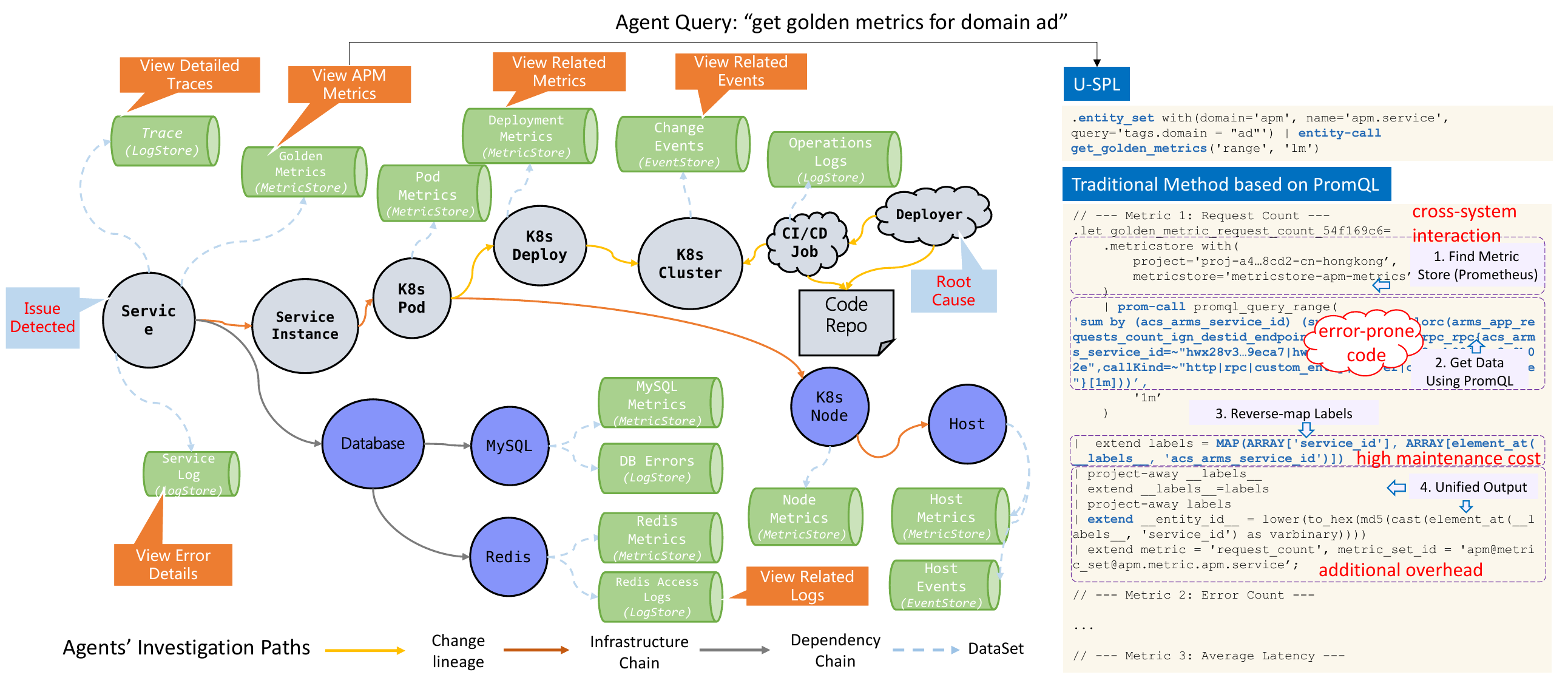}
  \caption{Comparison of query complexity between U-SPL and traditional methods. The top right depicts a concise U-SPL statement for retrieving golden metrics, while the bottom section illustrates the equivalent traditional approach, which involves complex PromQL scripting, manual label mapping, and cross-system interaction.}
\label{fig:spl-case}
\end{figure*}




By adopting an object-centric modeling paradigm, the entire data model can be represented as a knowledge-attached graph. However, the ultimate objective of a data model is to enable efficient and accurate querying. If traditional query mechanisms are still employed, it becomes necessary to generate different types of query statements (e.g., PromQL, SQL) to retrieve heterogeneous modalities of data (such as metrics and logs), and then join the results based on complex relationships defined in the graph. In agent-based scenarios, this approach introduces two major challenges: (1) the automated generation of query statements often suffers from a high error rate \cite{zhang2025promassistant}; and (2) in complex scenarios, join operations and schema mappings become highly intricate and error-prone.

To address these challenges, UModel proposes a unified, object-oriented query language called U-SPL. 
It abstracts away the underlying complexity of heterogeneous data sources to enable efficient retrieval and analysis of entity relationships. As illustrated in Figure~\ref{fig:spl-case}, the advantages of this approach are evident when performing a standard operations task in the ``Agents' Investigation Paths'', such as an agent querying the ``\textit{golden metrics for domain ad}''.
Using U-SPL, this intent is translated into a concise, two-line statement using high-level operators like \texttt{.entity\_set} and \texttt{entity-call}. In stark contrast, the traditional method relying on raw PromQL requires a multi-stage, error-prone process. The traditional approach forces the user to manually select the specific project and metric store (e.g., metricstore-apm-metrics), construct complex \texttt{promql\_query\_range} calls with explicit aggregations, and perform tedious post-processing steps such as reverse-mapping labels and converting entity IDs (e.g., \texttt{to\_hex(md5(...))}). By unifying these disparate steps, finding the store, retrieving data, and unifying output, into a single semantic abstraction, U-SPL significantly reduces maintenance costs and is more friendly for LLM-based agents.

\begin{table}[t]
\centering
\caption{Comparison of query types and target data.}
\label{tab:query_type_comparison}
\footnotesize
\resizebox{\linewidth}{!}{
\begin{tabular}{lll}
\toprule
\textbf{Query Type} & \textbf{Target} & \textbf{Examples} \\
\midrule
\textbf{MetaSearch} 
& Schema
& EntitySet's definition \\

\textbf{USearch} 
& Specific entity 
& Pod instance and related data \\

\textbf{GSearch}
& Relationship of entities 
& Call relationships \\
\bottomrule
\end{tabular}
}
\end{table}

As summarized in Table~\ref{tab:query_type_comparison}, U-SPL mainly consists of three components: \textit{MetaSearch}, \textit{USearch}, and \textit{GSearch}. MetaSearch is primarily used to query schema information related to EntitySets and is generally used less frequently. In the following sections, we provide a detailed introduction to USearch and GSearch.

\subsubsection{USearch}



USearch is a class of query syntax in U-SPL that focuses on concrete entity instances, such as specific services or pods. For example, USearch enables object-oriented queries to retrieve metrics associated with a given entity. However, raw multimodal data is typically characterized by large volume and low information density, making it difficult for agents to interpret effectively and therefore unsuitable for downstream AIOps tasks. To address this limitation, USearch introduces a series of mechanisms that transform retrieved data into a more structured and informative representation, thereby making UModel more agent-ready.

\noindent\paragraph{Search mechanism}
Searching textual data, such as log data, is a key problem in data querying. 
USearch supports multiple search mechanisms, including term-based search and phrase-based search. 
Phrase search automatically performs tokenization by default, 
different terms within a phrase are connected using the logical \texttt{OR}, 
and a match is satisfied if any term appears. 
In addition, search results are ranked according to relevance scores. 
Queries can be applied to all fields or restricted to specific fields. 
As shown in box~\ref{box:usearch}, this can be achieved by appending 
\texttt{query="xxxx"} after \texttt{\_\_type\_\_(.entity)} in the SPL syntax. 
Moreover, complex logical conditions such as \texttt{AND}, \texttt{OR}, and \texttt{NOT} 
are supported to further refine query results.

\begin{tcolorbox}[
  title={Examples of \textit{USearch} and \textit{GSearch}},
  colback=white,
  colframe=black,
  float,
  floatplacement=t,
  label={box:entityset}
]
\begin{lstlisting}[basicstyle=\ttfamily\scriptsize]
# USearch
## Multi-Keyword Search (OR by Default)
.entity with(query='kubernetes docker container', topk=50) 
## Filtering
.entity with(domain='apm', type='apm.service', query='production')
## Filtering with Fuzzy Matching
.entity with(domain='k8s', type='*.pod', query='error')
## Rank
.entity with(query='search_term', topk=100, groupTopk=10)
# GSearch
## graph-match
.topo | graph-match (s:"apm@apm.service" {__entity_id__: '123'})-[e]-(d) project s, e, d
## graph-call
.topo | graph-call getNeighborNodes('sequence', 1, [(:"app@app.operation" {__entity_id__: '73e'})])
## graph-call + cypher
.topo | graph-call cypher(`MATCH (s:``apm@apm.service`` {__entity_id__: '35a'})-[e]-(d) RETURN s, e, d`)
\end{lstlisting}
\label{box:usearch}
\end{tcolorbox}

\paragraph{Filtering mechanism}
Since UModel models datasets as entities, USearch enables filtering based on entity attributes. 
For example, \texttt{.entity with(domain=`apm')} filters out entities whose domain is not \texttt{apm}. 
Fuzzy matching is also supported by providing regular expressions for relevant fields. 
In addition, multi-field combined filtering is allowed. 
Several filtering examples are illustrated in box~\ref{box:usearch}.

\paragraph{Ranking mechanism}
Given the large volume of data, USearch may return numerous relevant results. 
Effectively ranking these results to surface the most relevant ones is therefore crucial. USearch adopts a multi-factor relevance scoring algorithm to compute the similarity between the query and each data record. 
For a single query term, the score is jointly determined by the term’s inverse document frequency (IDF) weight 
and the weight of the corresponding field. 
For multiple query terms, individual scores are accumulated, 
while both term frequency weights and field weights are taken into account. 
As shown in box~\ref{box:usearch}, parameters such as \texttt{topk} can be specified in the SPL syntax to control the output scale.

\subsubsection{GSearch}


The topological relationships among entities constitute a critical source of information in UModel, particularly for tasks such as RCA. To this end, UModel introduces GSearch to support graph-oriented queries. Notably, GSearch adopts the same query language, SPL, as other components in the system. GSearch mainly consists of two parts: \textit{graph-match} and \textit{graph-call}.

\paragraph{Graph-match}
The \textit{graph-match} component performs graph traversal queries based on path expressions and is designed to discover relational paths among entities. As illustrated in box~\ref{box:usearch}, a graph-match query must explicitly specify a starting node, including both its label and \texttt{entity\_id}. For intermediate nodes along the traversal path, only labels are allowed. Specifying node attributes is not supported. In addition, graph-match does not support cycle handling and therefore cannot process queries involving loops.

\paragraph{Graph-call}
\textit{Graph-call} serves as an invocation interface rather than a single fixed syntax. It acts as an execution entry point for graph algorithms or graph functions, including both built-in graph functions and Cypher~\cite{francis2018cypher} statements. Cypher is a general-purpose query language specifically designed for graph databases and is more powerful in advanced graph analysis scenarios, such as multi-hop traversal, conditional path queries, and aggregation operations. As shown in box~\ref{box:usearch}, graph-call can invoke functions such as \texttt{getNeighborNodes} to retrieve neighboring nodes and \texttt{getDirectRelations} to obtain direct relationships between nodes. Moreover, graph-call allows the direct embedding of Cypher queries for execution.

\subsection{Application Layer: MCP}
\label{sec:application-mcp}

To enable LLM-based agents to interact with UModel in a structured and controllable manner, we implement a dedicated Model Context Protocol (MCP) server at the application layer. MCP~\cite{hou2025model} provides a standardized interface between agents and external systems, allowing UModel to expose its unified representation of entities, datasets, relationships, topology, metrics, and logs as callable tools. Instead of requiring agents to directly access heterogeneous storage systems or proprietary backend APIs, the UModel MCP server abstracts these details into a set of operational interfaces. Through natural language interaction, operators can instruct agents to search entities, inspect topology, query telemetry, and perform diagnostic operations over the UModel-defined system context.

As shown in Table~\ref{tab:mcp-tools-hierarchy}, the UModel MCP server provides tools at both the IaaS and PaaS layers. The IaaS layer mainly wraps native observability and cloud service APIs, such as log service and cloud monitoring interfaces, and supports flexible low-level queries such as \texttt{execute\_spl(query)}. The PaaS layer further exposes higher-level, entity-centric tools, such as \texttt{search\_entity}, \texttt{get\_topology}, and \texttt{drill\_down}, which are closer to production operation workflows. This layered abstraction allows agents to reason over entities and dependencies before issuing detailed telemetry queries, reducing the burden of query construction while preserving access to raw operational data when needed. As a result, the integration of UModel, SPL, MCP, and LLM-based agents provides an extensible application layer for interactive and agent-centric intelligent operations.

\begin{table}[t]
\centering
\caption{Hierarchical organization of MCP server tools.}
\label{tab:mcp-tools-hierarchy}
\footnotesize
\resizebox{\linewidth}{!}{
\begin{tabular}{ll}
\toprule
\textbf{Layer / Tool} & \textbf{Description} \\
\midrule

\textbf{PaaS Layer} & \\
\midrule
\quad\textbf{Entity Management} & \\
\quad\quad umodel\_get\_entities & Retrieves entities from a specified entity set. \\
\quad\quad umodel\_get\_neighbor\_entities & Retrieves neighboring entities based on relationships. \\
\quad\quad umodel\_search\_entities & Searches entities using semantic or conditional criteria. \\
\midrule

\quad\textbf{Dataset Management} & \\
\quad\quad umodel\_list\_data\_set & Lists datasets associated with entity types. \\
\quad\quad umodel\_search\_entity\_set & Searches entity sets in the observability model. \\
\quad\quad umodel\_list\_related\_entity\_set & Retrieves logically related entity sets. \\
\midrule

\quad\textbf{Data Query} & \\
\quad\quad umodel\_get\_metrics & Queries time-series metric data. \\
\quad\quad umodel\_get\_golden\_metrics & Retrieves core performance indicators. \\
\quad\quad umodel\_get\_relation\_metrics & Queries metrics across entity relationships. \\
\quad\quad umodel\_get\_logs & Retrieves log data for entities. \\
\quad\quad umodel\_get\_events & Retrieves event data. \\
\quad\quad umodel\_get\_traces & Retrieves detailed distributed traces. \\
\quad\quad umodel\_search\_traces & Searches and filters trace summaries. \\
\quad\quad umodel\_get\_profiles & Retrieves performance profiling data. \\
\midrule

\textbf{IaaS Layer} & \\
\midrule
\quad cms\_text\_to\_promql & Translates natural language into PromQL queries. \\
\quad sls\_text\_to\_sql & Translates natural language into SQL queries. \\
\quad cms\_execute\_promql & Executes PromQL queries on metric services. \\
\quad sls\_execute\_sql & Executes SQL queries on log services. \\
\quad sls\_execute\_spl & Executes native SPL queries. \\
\quad sls\_list\_projects & Lists available log service projects. \\
\quad sls\_list\_logstores & Lists logstores within a project. \\
\bottomrule

\end{tabular}
}
\end{table}

%% file: deploymentandlessons.tex
UModel has been deployed in the production environment of Alibaba Cloud for over a year. It serves tens of thousands of workspaces and supports hundreds of cloud products and operational scenarios, including Container Services, APM, infrastructure, DevOps, real user monitoring, synthetic monitoring, log services, and cloud monitoring. In production, UModel unifies thousands of standardized configuration models, covering entities, datasets, relationships, and storage configurations across heterogeneous systems such as cloud-native services, databases, middleware, and network devices. The system sustains millions of entity and relationship operations per second and provides second-level query response for data retrieval at the scale of tens of millions.

\begin{figure}[t]
    \centering
    \includegraphics[width=0.8\linewidth]{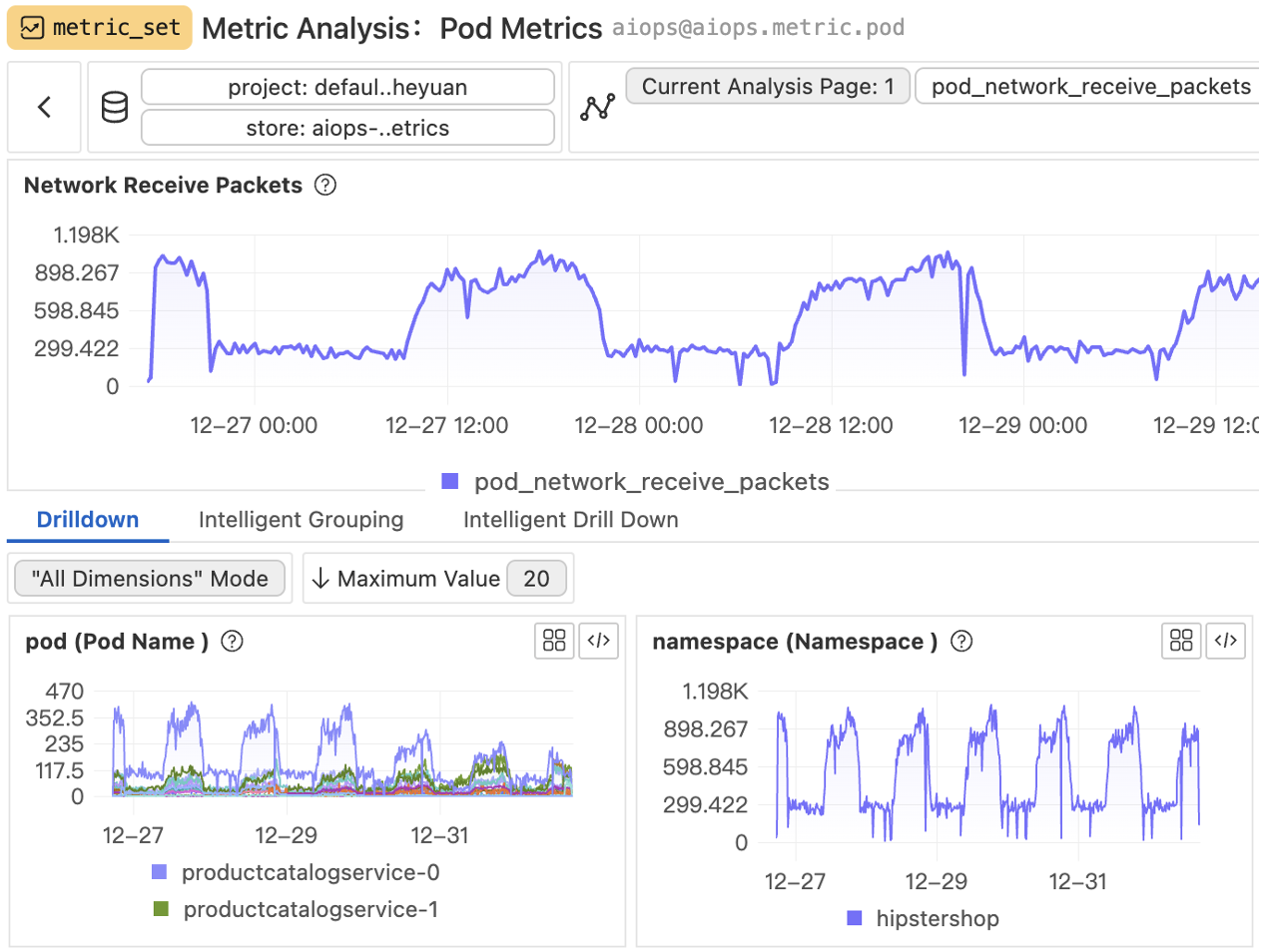}
    \caption{Drill-down in metricset explorer.}
    \label{fig:xiazuan}
\end{figure}


Several lessons were obtained from large-scale deployment. First, a unified data model should be complemented by task-oriented explorers. UModel provides both entity-oriented and metric-oriented explorers. In particular, the Metric Explorer supports anomaly detection, automated drill-down (illustrated in Figure~\ref{fig:xiazuan}), time-series clustering, label-based exploration, and root cause localization. These functions help operators identify suspicious signals without manually inspecting large numbers of metric curves. To maintain interactive performance, the explorer applies optimizations such as viewport visibility detection, query merging, and cascaded state update optimization.

Second, agent-centric root cause analysis requires more than direct access to queryable data. A straightforward design is to expose U-SPL directly to the RCA agent and let the LLM generate queries. However, this often leads to semantically incorrect queries, such as applying a \texttt{service\_id} filter to a node-level metric that only accepts \texttt{node\_id}. Moreover, raw query results may contain thousands of time-series points, which exceed the LLM context budget and degrade diagnosis reliability. To address this problem, we introduce a \textit{data-agent} layer in UModel that encapsulates U-SPL queries into domain-specific analysis tools, such as \textit{analyze\_golden\_metric(entity)} and \textit{analyze\_error\_log(entity)}. The main-agent (such as RCA agent) decides what to investigate, while the data-agent determines how to query, execute, and summarize the results. This separation shields the reasoning agent from query syntax and optimization details, reduces hallucination, and allows the data layer to evolve independently. The data-agent further compresses raw telemetry into concise observations, reducing average context consumption by over 80\% while improving diagnostic accuracy.

%% file: realworlddeployment.tex


    
    

We assess the effectiveness of UModel by investigating the following research questions:

\begin{itemize}[leftmargin=*]
    \item \textbf{RQ1:} How agent-ready is UModel compared with traditional data models for RCA tasks?
    
    \item \textbf{RQ2:} How does each component of UModel contribute to RCA performance?
    
    \item \textbf{RQ3:} In real-world production scenarios, do UModel's IaaS-level tools or PaaS-level tools provide greater effectiveness for RCA?
    
\end{itemize}

\subsection{RQ1: Traditional Data Model vs. UModel}


The effectiveness of RCA largely depends on whether the data model and its associated analytical interfaces can expose key signals and actionable clues to the agent. To demonstrate that UModel is more agent-ready than traditional data models, we evaluate its performance in an agent-based setting.




\subsubsection{Experimental details}

Our dataset comprises 138 faults injected using ChaosMesh, covering node-level, pod-level, and service-level failures. To comprehensively evaluate RCA performance, we employ two categories of metrics: quality metrics and ranking metrics. For quality evaluation, we report location accuracy with redundancy penalty (LA), type accuracy (TA), and reasoning score (RS). LA measures whether the agent correctly identifies the root-cause locations while penalizing redundant or incorrect predictions. Specifically, $L_c$ denotes the number of correctly identified root-cause locations, $L_i$ denotes the number of incorrectly identified locations, $L_t$ denotes the total number of ground-truth root causes, and $\sigma$ is a penalty coefficient, which is set to 0.05 by default. RS evaluates the interpretability of the agent's reasoning process by checking whether its reasoning chain contains predefined annotated keywords. For ranking-based evaluation, we report Top-$k$ location accuracy, which measures whether the ground-truth root cause appears within the top-$k$ ranked predictions.

\begin{equation}
\small
{
    \mathrm{LA} = \frac{L_c - \sigma \times L_i}{L_t}
}
\label{eq:la}
\end{equation}


For comparison, we adopt a ReAct-based agent~\cite{react2023} and connect it separately to traditional data models and to UModel. Under the traditional data model setting, the agent is equipped with tools for data querying, such as \texttt{get\_metrics}, log anomaly detection, such as \texttt{log\_anomaly\_detection}, and system information retrieval. Under the UModel setting, the agent interacts with an MCP server that exposes a suite of IaaS- and PaaS-level tools, including SPL statement execution, as detailed in Table~\ref{tab:mcp-tools-hierarchy}.

\begin{table}[t]
\centering
\caption{Comparison of RCA performance between traditional data model and UModel. The LLM backbone we use is Qwen3-max.}
\label{tab:rca_comparison}
\resizebox{0.95\linewidth}{!}{
\begin{tabular}{>{\bfseries}lccc}
\toprule
\textbf{Metric} & \textbf{Traditional Data Model} & \textbf{UModel} & \textbf{Improvement} \\
\midrule
\multicolumn{4}{l}{\textit{Quality Metrics}} \\
\hline
OS & 49.70 & \textbf{55.45} & {\scriptsize\color{BrickRed}(↑5.75)} \\
TA            & 36.11 & \textbf{40.74} & {\scriptsize\color{BrickRed}(↑4.63)} \\
LA          & 55.96 & \textbf{64.08} & {\scriptsize\color{BrickRed}(↑8.12)} \\
RS            & 48.56 & \textbf{58.53} & {\scriptsize\color{BrickRed}(↑9.97)} \\
\hline
\multicolumn{4}{l}{\textit{Ranking Metrics}} \\
\hline
Top-1 Acc & 68.12 & \textbf{74.64} & {\scriptsize\color{BrickRed}(↑6.52)} \\
Top-3 Acc & 71.74 & \textbf{78.99} & {\scriptsize\color{BrickRed}(↑7.25)} \\
\bottomrule
\end{tabular}
}
\end{table}

\subsubsection{Overall Result}
The overall results are summarized in Table~\ref{tab:rca_comparison}. As shown, UModel consistently outperforms the traditional data model across multiple metrics in agent-based RCA tasks, achieving an 8\% improvement in location accuracy and a 10\% increase in reasoning score. The observed gains in TA and RS indicate that, by leveraging \textit{MetaSearch} (e.g., querying event-related information), UModel enables agents to more effectively retrieve relevant keywords, thereby improving fault type classification accuracy and enhancing reasoning interpretability. Meanwhile, the improvements in LARP and Top-$k$ Acc suggest that UModel, through \textit{GSearch} (e.g., exploring relationships among pods), more effectively captures complex dependency structures, leading to more accurate root cause localization.

It is worth noting that the 8.12\% improvement in the overall results represents a conservative estimate. The primary reason is that, initially, to ensure a fair comparison, we require both the baseline agent and the UModel-based agent to autonomously compose query languages to retrieve observability data, including metrics, logs, and traces. However, we observe that for the baseline, the precision of directly generating PromQL is less than 5\%, resulting in a failure to obtain valid data. This limitation is corroborated by recent literature, for instance, previous work indicates that the accuracy of generating PromQL using GPT-4-Turbo is only about 2.6\%~\cite{zhang2025promassistant}. This accuracy is measured by \textit{QueryAcc}, defined as the percentage of correct queries out of all generated queries, where a query is considered correct if it matches the ground truth or is a semantically equivalent implementation.

To ensure the evaluation proceeds smoothly, we modify the baseline setup by pre-loading all necessary observability data into a DataFrame, thereby granting the baseline agent 100\% data accessibility. Consequently, the reported improvement reflects only the gains from UModel's reasoning and structural design, \textbf{excluding} the substantial benefits regarding U-SPL's object-centric nature. In practice, the object-centric design of U-SPL drastically reduces the hallucination rate and enhances the agent's success in data retrieval, which constitutes a significant advantage over the baseline.

\subsection{RQ2: Ablation Study}



As described in Section~\ref{sec:design}, the upper layer of UModel mainly consists of three components: MetaSearch, USearch, and GSearch. Since MetaSearch is primarily used to query EntitySet information and is generally not involved in the evaluated workflows, we do not consider it in this ablation study. Instead, we independently remove USearch and GSearch to evaluate their respective contributions. It is worth noting that some query indicators and log traces included in USearch are also available in traditional data models; therefore, these aspects are not the focus of our analysis. In this study, we only disable the newly introduced functionality in UModel, specifically the capability of USearch to query event-related data. For GSearch, we remove the functionality related to querying inter-entity relationships.

The results presented in Table~\ref{tab:ablation} indicate that GSearch has a substantial impact on location-related metrics, as it effectively identifies relationships among different entities. In contrast, USearch enables rapid retrieval of relevant entity information---especially data related to critical events---and therefore affects all evaluation metrics to varying degrees.

\begin{table*}[t]
\centering
\caption{Comparison between the IaaS query layer and the PaaS semantic layer across three evaluation sets. 
All OS, LA, and TA values are reported on a 0--100 scale.}
\label{tab:paas-vs-iaas}
\resizebox{0.8\textwidth}{!}{
\begin{tabular}{ccccccccccc}
\toprule
Set & Framework
& IaaS OS & PaaS OS & $\Delta$ OS
& IaaS LA & PaaS LA & $\Delta$ LA
& IaaS TA & PaaS TA & $\Delta$ TA \\
\midrule
Full Paired & Multi-Debate
& 32.50 & 46.47 & \color{BrickRed}(↑13.97)
& 27.00 & 58.10 & \color{BrickRed}(↑31.10)
& 18.50 & 19.00 & \color{BrickRed}(↑0.50) \\

Full Paired & OpenClaw
& 34.90 & 40.94 & \color{BrickRed}(↑6.04)
& 30.90 & 45.60 & \color{BrickRed}(↑14.70)
& 23.20 & 19.10 & \color{ForestGreen}(↓4.10) \\

Full Paired & Plan-Execute
& 24.38 & 28.25 & \color{BrickRed}(↑3.87)
& 11.10 & 21.20 & \color{BrickRed}(↑10.10)
& 11.30 & 10.80 & \color{ForestGreen}(↓0.50) \\

Full Paired & ReAct
& 29.02 & 43.98 & \color{BrickRed}(↑14.96)
& 18.20 & 51.50 & \color{BrickRed}(↑33.30)
& 21.60 & 21.20 & \color{ForestGreen}(↓0.40) \\

Full Paired & Reflexion
& 27.71 & 35.32 & \color{BrickRed}(↑7.61)
& 15.70 & 33.30 & \color{BrickRed}(↑17.60)
& 20.30 & 20.50 & \color{BrickRed}(↑0.20) \\

Full Paired & \textbf{Pool}
& \textbf{29.70} & \textbf{38.99} & \textbf{\color{BrickRed}(↑9.29)}
& \textbf{20.60} & \textbf{42.00} & \textbf{\color{BrickRed}(↑21.40)}
& \textbf{19.00} & \textbf{18.10} & \textbf{\color{ForestGreen}(↓0.90)} \\
\midrule

Stratified Main & Multi-Debate
& 28.34 & 53.45 & \color{BrickRed}(↑25.11)
& 18.20 & 68.20 & \color{BrickRed}(↑50.00)
& 12.50 & 28.60 & \color{BrickRed}(↑16.10) \\

Stratified Main & OpenClaw
& 44.15 & 45.70 & \color{BrickRed}(↑1.55)
& 50.00 & 55.00 & \color{BrickRed}(↑5.00)
& 27.50 & 25.00 & \color{ForestGreen}(↓2.50) \\

Stratified Main & Plan-Execute
& 22.11 & 28.79 & \color{BrickRed}(↑6.68)
& 3.60 & 21.40 & \color{BrickRed}(↑17.80)
& 14.30 & 14.30 & 0.00 \\

Stratified Main & ReAct
& 27.52 & 47.47 & \color{BrickRed}(↑19.95)
& 14.30 & 57.10 & \color{BrickRed}(↑42.80)
& 20.60 & 26.40 & \color{BrickRed}(↑5.80) \\

Stratified Main & Reflexion
& 26.18 & 34.51 & \color{BrickRed}(↑8.33)
& 13.30 & 33.30 & \color{BrickRed}(↑20.00)
& 16.60 & 18.50 & \color{BrickRed}(↑1.90) \\

Stratified Main & \textbf{Pool}
& \textbf{29.66} & \textbf{41.99} & \textbf{\color{BrickRed}(↑12.33)}
& \textbf{19.90} & \textbf{47.00} & \textbf{\color{BrickRed}(↑27.10)}
& \textbf{18.30} & \textbf{22.60} & \textbf{\color{BrickRed}(↑4.30)} \\
\midrule

Filtered Effective & ReAct
& 25.80 & 48.74 & \color{BrickRed}(↑22.94)
& 13.30 & 56.70 & \color{BrickRed}(↑43.40)
& 21.80 & 28.00 & \color{BrickRed}(↑6.20) \\

Filtered Effective & OpenClaw
& 22.48 & 44.40 & \color{BrickRed}(↑21.92)
& 20.00 & 56.70 & \color{BrickRed}(↑36.70)
& 17.50 & 16.00 & \color{ForestGreen}(↓1.50) \\

Filtered Effective & Multi-Debate
& 27.06 & 45.08 & \color{BrickRed}(↑18.02)
& 14.80 & 58.30 & \color{BrickRed}(↑43.50)
& 19.70 & 15.80 & \color{ForestGreen}(↓3.90) \\

Filtered Effective & Reflexion
& 25.92 & 28.81 & \color{BrickRed}(↑2.89)
& 13.30 & 23.30 & \color{BrickRed}(↑10.00)
& 16.40 & 14.50 & \color{ForestGreen}(↓1.90) \\

Filtered Effective & Plan-Execute
& 25.37 & 23.13 & \color{ForestGreen}(↓2.24)
& 13.30 & 10.00 & \color{ForestGreen}(↓3.30)
& 13.30 & 16.70 & \color{BrickRed}(↑3.40) \\

Filtered Effective & \textbf{Pool}
& \textbf{25.29} & \textbf{37.74} & \textbf{\color{BrickRed}(↑12.45)}
& \textbf{15.00} & \textbf{40.30} & \textbf{\color{BrickRed}(↑25.30)}
& \textbf{17.70} & \textbf{18.20} & \textbf{\color{BrickRed}(↑0.50)} \\
\bottomrule
\end{tabular}
}
\end{table*}

\begin{table}[t]
\centering
\small
\caption{Ablation study results (\%). U = USearch, G = GSearch.}
\setlength{\tabcolsep}{5pt}
\resizebox{0.95\linewidth}{!}{
\begin{tabular}{lcccc}
\toprule
\textbf{Setting} & \textbf{TA} & \textbf{LARP} & \textbf{Top-1} & \textbf{Top-3} \\
\hline
G\&U   & 36.96 & 66.34 & 68.84 & 74.64 \\
w/o G  & 30.43 {\scriptsize\color{ForestGreen}(↓6.53)}
       & 62.39 {\scriptsize\color{ForestGreen}(↓3.95)}
       & 68.84 {\scriptsize\color{ForestGreen}(↓0.00)}
       & 74.64 {\scriptsize\color{ForestGreen}(↓0.00)} \\
w/o U  & 22.46 {\scriptsize\color{ForestGreen}(↓14.50)}
       & 55.48 {\scriptsize\color{ForestGreen}(↓10.86)}
       & 64.49 {\scriptsize\color{ForestGreen}(↓4.35)}
       & 71.01 {\scriptsize\color{ForestGreen}(↓3.63)} \\
w/o G\&U & 20.29 {\scriptsize\color{ForestGreen}(↓16.67)}
         & 57.50 {\scriptsize\color{ForestGreen}(↓8.84)}
         & 64.83 {\scriptsize\color{ForestGreen}(↓4.01)}
         & 69.83 {\scriptsize\color{ForestGreen}(↓4.81)} \\
\bottomrule
\end{tabular}
}
\label{tab:ablation}
\end{table}

\subsection{RQ3: IaaS vs. PaaS in Real-World Production Scenarios}

\subsubsection{Experimental Setting}

To examine which tool abstraction is more effective for RCA in real-world production scenarios, we compare two tool-layer designs under the same conditions. We evaluate both settings on five agent frameworks: ReAct \cite{react2023}, OpenClaw \cite{openclaw2026}, Multi-Debate \cite{chan2024chateval}, Reflexion \cite{shinn2023reflexion}, and Plan-Execute \cite{wang2023plan}. For each paired comparison, we keep the LLM configuration, prompts, maximum interaction budget, scoring protocol, and case set fixed. We report the overall score $\mathrm{OS} = 0.4\mathrm{TA} + 0.3\mathrm{LA} + 0.3\mathrm{RS}$, together with LA and TA. These metrics are similar to those used in RQ1. All metrics are reported on a 0--100 scale.


To reduce sensitivity to a single benchmark split, we conduct the comparison on three evaluation sets. The \textbf{Full Paired Set} contains all valid paired trials from the full benchmark. The \textbf{Stratified Main Set} is the pre-registered 30-case stratified subset. The \textbf{Filtered Effective Set} removes floor-effect cases where both tool layers fail to obtain discriminative evidence.

\subsubsection{Results}

Table~\ref{tab:paas-vs-iaas} shows that the PaaS semantic layer consistently improves the pooled overall score (OS) over the IaaS query layer across all three evaluation sets. The pooled OS increases by 9.29 points on the Full Paired Set, 12.33 points on the Stratified Main Set, and 12.45 points on the Filtered Effective Set.

The improvement is mainly driven by location accuracy. Across the three pooled results, LA improves by 21.40, 27.10, and 25.30 points, respectively. By contrast, TA changes only slightly, with differences of -0.90, +4.30, and +0.50 points. This indicates that the PaaS semantic layer primarily helps agents locate the correct root-cause entity, rather than substantially changing their ability to infer the fault type.

This result aligns with the challenges of real-world production RCA. In the IaaS setting, agents must directly compose SPL queries and map raw resource identifiers to operational entities. In the PaaS setting, normalized entity identifiers and topology-aware retrieval reduce this burden, allowing agents to focus more on causal reasoning. This abstraction is especially useful when failures span multiple infrastructure and application layers.

At the framework level, ReAct and Multi-Debate show the most pronounced improvements, suggesting that iterative exploration benefits strongly from semantic entity search and topology expansion. Reflexion obtains moderate gains, while Plan-Execute is less stable and shows a negative OS gain on the Filtered Effective Set. Overall, the results suggest that, in real-world production-oriented RCA scenarios, PaaS-level semantic tools provide a more effective abstraction than direct IaaS-level SPL access.



\subsection{Case Study}
\label{subsec:case-study}

\begin{figure}
    \centering
    \includegraphics[width=\linewidth]{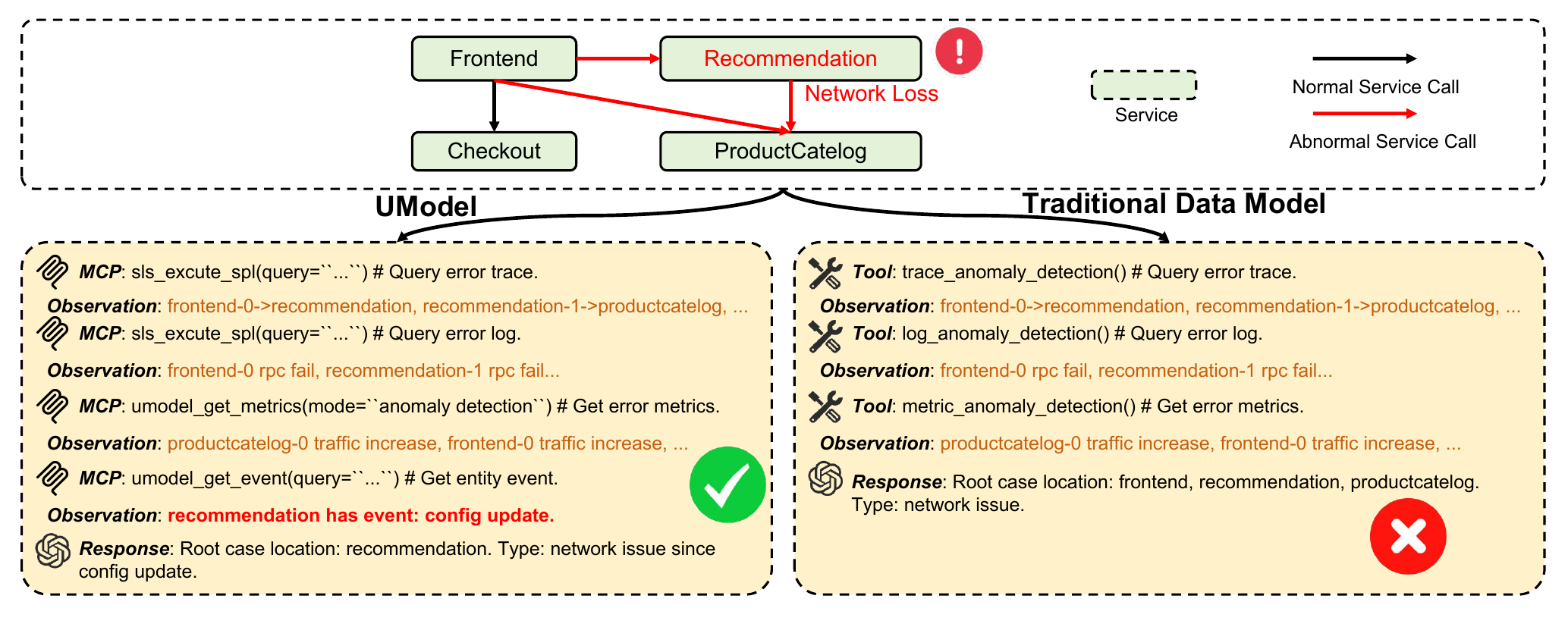}
    \caption{Agent call chains of UModel and traditional data model.}
    \label{fig:case1}
\end{figure}

To further illustrate why UModel achieves superior RCA performance compared to traditional data-driven models, we conduct a case study based on two representative categories of incidents. In these cases, USearch and GSearch respectively play critical roles in uncovering key information about the root causes that traditional data models fail to capture. We describe these cases in detail below.

\subsubsection{Case 1: USearch}

As shown in Figure~\ref{fig:case1}, the entire system experienced a network loss failure due to an incorrect configuration update in the recommendation service. Since the recommendation service occupies a central and critical position in the overall call topology, its failure rapidly propagated to downstream components, including the frontend and product catalog services. 

When performing RCA using traditional data models, a variety of anomaly detection tools—such as trace anomaly detection, log anomaly detection, and time-series anomaly detection—are typically invoked. These tools collectively generate a large volume of error signals across multiple services. However, the resulting information is often noisy, unstructured, and lacks prioritization, making it difficult to identify the true root cause. Consequently, the RCA agent tends to treat many abnormal services as potential root causes and includes the majority of them in the final diagnosis results.

In contrast, UModel adopts an object-centric modeling paradigm that explicitly represents entities and their associated events. Configuration updates, as first-class events, can be easily incorporated into UModel through visualization interfaces or other integration mechanisms, and can be queried via MCP. Once the agent is aware that a configuration change occurred in the recommendation service during the failure window, it can accurately localize the root cause.

\subsubsection{Case 2: GSearch}

\begin{figure}
    \centering
    \includegraphics[width=0.7\linewidth]{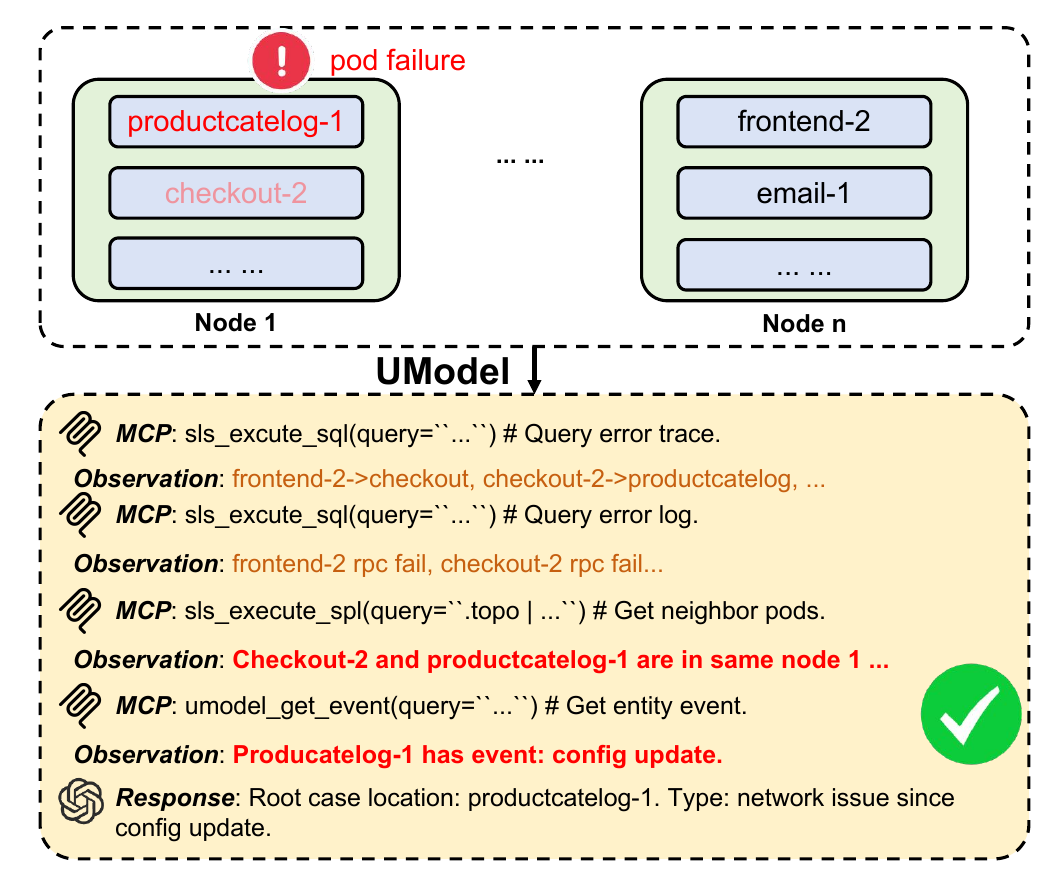}
    \caption{Agent call chain of UModel.}
    \label{fig:case2}
\end{figure}


As illustrated in Figure~\ref{fig:case2}, when \texttt{productcatalog-1} experienced a failure due to an erroneous change, other services co-located on the same node were also impacted because they shared underlying resources such as CPU, memory, and network bandwidth. In particular, \texttt{checkout-2}, which was deployed on the same node, was highly susceptible to this disturbance. As a result, multiple pods triggered alerts during the anomaly detection phase. Moreover, since services differ in their criticality and sensitivity, the indirectly affected pod (\texttt{checkout-2}) may generate a larger number of alerts or exhibit more severe anomaly signals than the actual root cause. In such scenarios, traditional data-driven models tend to bias the RCA agent toward selecting \texttt{checkout-2} as the root cause, as it appears to be the most anomalous component.

In contrast, UModel enables the agent to explicitly query various structural and dependency relationships using SPL statements and to retrieve event-level information associated with entities. If the agent discovers that \texttt{productcatelog-1} on the same node has a relevant event during the failure window, it can infer with high confidence that this node—and the change in \texttt{productcatelog-1}—is the true root cause.





%% file: RelatedWork.tex
Modern AIOps builds on observability pipelines that transform running systems into machine-consumable signals. The canonical triad of \emph{metrics, logs, and traces} has been institutionalized through distributed tracing infrastructures and standardization efforts that unify context propagation, collection protocols, and semantic conventions across platforms and languages~\citep{ding2024rd,zheng2024mumon,chen2024eagle}. In production environments, metrics are typically scraped and aggregated by time-series monitoring systems~\citep{prometheus,mehner2024ipd,zhang2025achieving}, while logs rely on standardized transport and message formats~\citep{rfc5424_syslog}. Low-overhead instrumentation is increasingly enabled at the kernel and dataplane levels: packet- and syscall-level observability originates from BPF~\citep{bpf1993}, and eBPF extends this model to programmable in-kernel telemetry for profiling, networking, and security observability at scale~\citep{ebpf_runtime_2024}. In networking, flow-level telemetry~\citep{rfc3954_netflow9,rfc3176_sflow,wang2025hawkeye} and programmable dataplanes~\citep{p4_2014} further broaden the observability surface, linking application- and network-level measurements.

Building on these signals, substantial work addresses core AIOps tasks such as anomaly detection and RCA. For metrics, approaches range from probabilistic and deep learning models for time-series anomaly detection~\citep{omnianomaly2019,wang2024revisiting} to attention-based sequence models~\citep{anomaly_transformer2021,cui2025tshape}. Log-based methods commonly parse raw logs into structured templates~\citep{drain2017} and apply sequential or representation learning models for detection and diagnosis~\citep{deeplog2017,logbert2022}. RCA in distributed and microservice systems frequently leverages graph-based and causal reasoning to localize faulty services or indicators, including graph-based localization~\citep{microrca2020,xie2023point,sundara2023global}, intra-service causal analysis~\citep{microcause2020,wang2025towards}, and propagation-chain analysis optimized for efficiency~\citep{microhecl2021,liu2025skeletonhunter,yang2025skynet}. Recent systems increasingly exploit multimodal fusion of \emph{metrics, logs, and traces} to improve diagnostic robustness, as demonstrated in multimodal microservice failure diagnosis and tracing-centric learning frameworks~\citep{diagfusion2023,nedelkoski2019,nie2025dest}. More recently, agentic and LLM-based AIOps has emerged as an interface layer for tool orchestration, reasoning over heterogeneous signals, and automating runbook-style operational workflows~\citep{react2023,toolformer2023,llm4aiops_survey2025}.

%% file: conclusion.tex

In this paper, we identify that a key bottleneck in modern agent-driven AIOps lies in the fragmentation and limited semantics of traditional observability data organization. To address this, we propose \textit{UModel}, an agent-ready data model that unifies heterogeneous operational data into an object-centric graph, linking entities, telemetry, and remediation tools. The proposed U-SPL further enables active system exploration, allowing LLM-based agents to retrieve topology-aware context and perform multi-hop reasoning without rigid pre-processing. Deployment in a large-scale cloud production environment demonstrates that UModel sustains industrial-grade performance at 10 million operations per second and significantly improves RCA effectiveness.

%% file: appendix.tex
\subsection{Visualized Model Building}
\label{sec:visualizedmodelbuild}

Constructing a data model from large volumes of raw multimodal data is labor-intensive. Therefore, ease of construction is a prerequisite for the rapid adoption of an observable data model. Through continuous iteration, UModel has streamlined the construction process into a small number of intuitive steps:

\begin{enumerate}[leftmargin=*]
    \item \textbf{Identification of entities and datasets.}  
    Entity types vary across systems. For example, in Kubernetes environments, entities typically include \texttt{service}, \texttt{pod}, and \texttt{node}, whereas in data center environments, entities correspond to nodes at different hierarchical levels.
    
    \item \textbf{Construction of entities and datasets.}  
    Based on the identified entities and datasets, schema files are generated from predefined templates. These schema files are then imported into the system to instantiate the corresponding entity sets and dataset sets.
    
    \item \textbf{Construction of links.}  
    Relationships exist both between different entity sets and between entity sets and datasets. These relationships are explicitly modeled as links, which are also defined through schema files. For example, in Kubernetes, \texttt{service} entity \emph{contains} \texttt{pod}.
    
    \item \textbf{Data ingestion.}  
    Once the UModel is fully defined, real data can be ingested into the system, enabling the complete UModel pipeline to operate.
    
    \item \textbf{Data validation.}  
    Finally, simple SPL queries are executed to verify the correctness of the constructed UModel.
\end{enumerate}

\begin{figure}[t]
    \centering
    \includegraphics[width=0.85\linewidth]{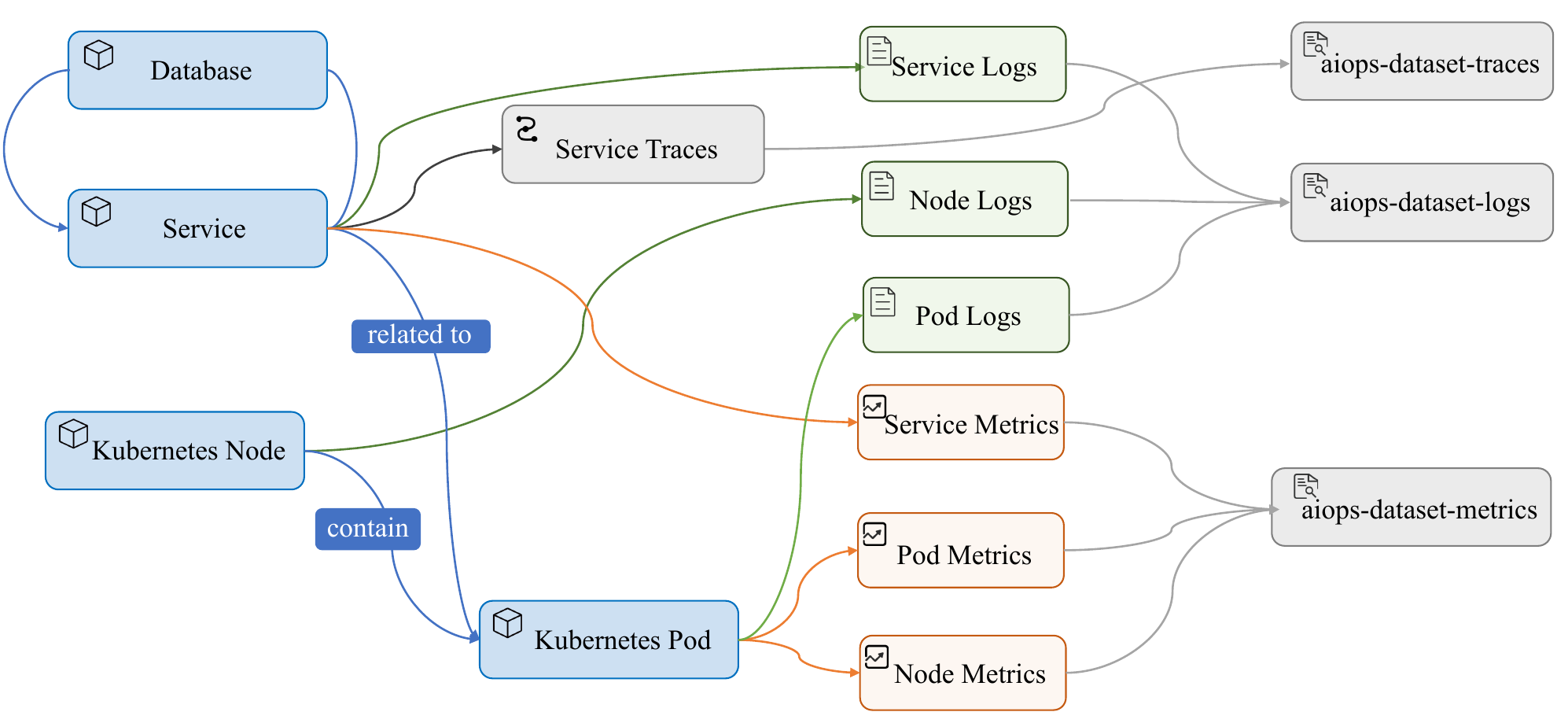}
    \caption{Viusalized model construction.}
    \label{fig:visualedit}
\end{figure}

Among these steps, the primary manual effort lies in defining entities and links via schema files. To further reduce this overhead, UModel introduces a visualized editing mechanism, as illustrated in Figure~\ref{fig:visualedit}. Entity sets and links can be created, modified, or removed directly through a web-based interface by clicking the ``+'' button, without manually writing YAML schema files. Attributes can be interactively added by selecting the corresponding entity or link, significantly improving usability and construction efficiency.

\subsection{QA Tasks in AIOps}
\label{sec:qa}

To illustrate the advantages of the \textbf{UModel SPL} mechanism over traditional operational workflows, we design three representative query scenarios covering data aggregation, topology reasoning, and joint data–knowledge analysis.

\textbf{Q1 (Data Query)} focuses on retrieving aggregated metrics from multiple Pods. In traditional approaches, operators must query each Pod’s metric data individually and perform aggregation manually at the application or scripting layer. In contrast, \textbf{UModel} enables batch entity selection and server-side aggregation within a single \textbf{SPL} query, significantly simplifying the workflow and reducing redundant data retrieval.

\textbf{Q2 (Topology Knowledge Reasoning)} targets the extraction of a multi-hop dependency subgraph starting from a faulty Pod. Conventionally, this requires multiple heterogeneous steps, including collecting ownership information, merging dependency relations from traces and logs, constructing a unified relationship graph, and iteratively expanding neighbors using multi-join or BFS-like logic. \textbf{UModel} directly leverages its built-in topology graph and graph-call interface, allowing the entire multi-hop neighborhood to be retrieved through a single declarative graph query, eliminating manual graph construction and traversal.

\textbf{Q3 (Data and Knowledge Query)} combines structural knowledge with metric retrieval by querying metrics of Pods deployed on the same node as a target Pod. The traditional process involves first identifying the hosting node, then enumerating peer Pods, and finally querying metrics for each Pod. \textbf{UModel} integrates topology traversal and metric querying into a unified pipeline: topology relations are used to derive peer Pod identifiers, which are immediately passed to metric retrieval operations. This tightly coupled data–knowledge interaction reduces procedural complexity and improves query efficiency.

As shown in Table~\ref{tab:QA}, we compare traditional methods with the \textbf{UMODEL SPL} method.

\begin{table*}[t]
\newcolumntype{P}[1]{>{\raggedright\arraybackslash}p{#1}}
\centering
\caption{Comparison between traditional methods and UModel SPL method.}
\label{tab:QA}
\footnotesize
\renewcommand{\arraystretch}{1.25}
\begin{tabular}{c|p{5cm}|p{4cm}|p{4cm}}
\toprule
\multirow{2}{*}{\textbf{Question}} & \multirow{2}{*}{\textbf{Traditional Methods}} & \multicolumn{2}{c}{\textbf{UModel SPL}} \\ \cline{3-4}
 &  & \textbf{SPL Sample} & \textbf{Result/Explanation} \\ 
\hline
Q1 (Data)
&
\begin{tabular}[t]{@{}l@{}}
1) Query \{pod1\} \{metric\} data; \\
2) Query \{pod2\} \{metric\} data; \\
3) Query \{pod3\} \{metric\} data; \\
4) Aggregate and return.
\end{tabular}
&
\begin{lstlisting}[
  basicstyle=\normalfont,
  breaklines=true,
  aboveskip=-7pt,
  belowskip=0pt,
  xleftmargin=0pt
]
.entity_set(with(domain='aiops', name='aiops.pod', ids=[id1,id2,...])) | entity-call get_metric('aiops','aiops.metric.pod','{metric}','range','',aggregate=true)
\end{lstlisting}
&
Aggregates multi-pod metric values in one query and returns unified results. \\
\hline

Q2 (Knowledge) 
& 
\begin{tabular}[t]{@{}l@{}}
1) Get ownership (node/service).\\
2) Collect depth from trace/logs.\\
3) Merge into a unified dep graph/table.\\
4) Expand to 4 hops (BFS / multi-join).\\
5) Summarize and filter related entities.
\end{tabular}
&
\begin{lstlisting}[
  basicstyle=\normalfont,
  breaklines=true,
  aboveskip=-7pt,
  belowskip=0pt,
  xleftmargin=0pt
]
.topo | graph-call cypher(`MATCH (s:`aiops@aiops.pod` {__entity_id__:'id'})-[e]-(d)-[f]-(g)-[h]-(j)-[k]-(l)RETURN s, d, g, j, l`)
\end{lstlisting}
&
1-query 4-hop subgraph extraction; No manual multi-join/BFS.\\
\hline

Q3 (Data+Knowledge)
& 
\begin{tabular}[t]{@{}l@{}}
1) Resolve the Pod's node.\\
2) List all Pods on that node.\\
3) Query \{metric\} for these Pods.
\end{tabular}
& 
\begin{lstlisting}[
  basicstyle=\normalfont,
  breaklines=true,
  aboveskip=-7pt,
  belowskip=0pt,
  xleftmargin=0pt
]
1) .topo | graph-call cypher( MATCH (s:aiops@aiops.pod {__entity_id__:'id'})-[e]-(d)-[f]-(g) RETURN g.__entity_id__ AS entity_ids )} 
2) .entity_set with(domain='aiops', name='aiops.node', ids=['entity_ids']) | entity-call get_metric('aiops','aiops.metric.node','{metric}','range','',aggregate=false)
\end{lstlisting}
&
Topology query first finds peer Pods on the same node, then batch metric retrieval is executed in one flow without manual joins.\\
\bottomrule
\end{tabular}
\end{table*}

\subsection{Core Concept of UModel}
\label{sec:concept}


Here we present several schema examples of core concepts, as illustrated in the boxes below (box C.1, C.2, C.3 and C.4). 

\begin{tcolorbox}[
  title={C.1: Entity Set Schema},
  colback=white,
  colframe=black,
  float,
  floatplacement=htbp,
  label={box:entityset}
]
\begin{lstlisting}[basicstyle=\ttfamily\scriptsize]
kind: entity_set
metadata:
  name: "aiops.service"
  display_name:
    zh_cn: "service"
  domain: aiops
spec:
  fields:
    - name: service
      type: string
      example: "adservice"
  primary_key_fields:
    - "service"
  name_fields:
    - "service"
\end{lstlisting}
\end{tcolorbox}

\begin{tcolorbox}[title={C.2: Entity Set Link Schema}, colback=white, colframe=black, float, floatplacement=htbp]
\begin{lstlisting}[basicstyle=\ttfamily\scriptsize]
kind: entity_set_link
metadata:
  name: "aiops.service_calls_aiops.service" 
  domain: aiops
spec:
  left_entity_set: 
    domain: aiops
    name: aiops.service
  right_entity_set: 
    domain: aiops
    name: aiops.service
  link_type: calls
  link_mapping:
    match_type: exact
    fields:
      - left: "service"
        right: "service"
\end{lstlisting}
\end{tcolorbox}

\begin{tcolorbox}[
  title={C.3: Metric Set Schema},
  colback=white,
  colframe=black,
  float,
  floatplacement=htbp,
  label={box:entityset}
]
\begin{lstlisting}[basicstyle=\ttfamily\scriptsize]
kind: metric_set
metadata:
  name: "aiops.metric.service"
  display_name:
    zh_cn: "service metric"
  domain: aiops
spec:
  labels:
    dynamic: false
    keys:
      - name: service
        type: string
        example: "adservice"
  metrics:
    - name: "request"
      display_name:
        zh_cn: "request number"
      golden_metric: true
      generator: 'request'
      aggregator: 'sum'
\end{lstlisting}
\end{tcolorbox}

\begin{table*}[htbp]
\centering
\caption{Data Set Configuration Specification}
\label{tab:unified_set_config}
\footnotesize
\resizebox{0.8\linewidth}{!}{
\begin{tabular}{lll}
\toprule
\textbf{Configuration Item} & \textbf{Description} & \textbf{Notes} \\
\midrule

\multicolumn{3}{l}{\textbf{Metric Set Configuration}} \\
\midrule
\texttt{metadata.name} 
& Metric set identifier 
& Format: \texttt{\{domain\}.metric.\{entity\_name\}} \\

\texttt{spec.labels} 
& Label definitions 
& Used for filtering and grouping (e.g., \texttt{service}, \texttt{region}) \\

\texttt{spec.metrics} 
& Metric definitions 
& Concrete monitoring metrics (e.g., \texttt{request}, \texttt{error}) \\

\midrule
\multicolumn{3}{l}{\textbf{Log Set Configuration}} \\
\midrule
\texttt{metadata.name} 
& Log set identifier 
& Format: \texttt{\{domain\}.log.\{entity\_name\}} \\

\texttt{spec.time\_field} 
& Time field 
& Typically set to \texttt{\_\_time\_\_} \\

\texttt{spec.fields} 
& Field definitions 
& Includes all log fields \\

\midrule
\multicolumn{3}{l}{\textbf{Trace Set Configuration}} \\
\midrule
\texttt{metadata.name} 
& Trace set identifier 
& Format: \texttt{\{domain\}.trace.\{entity\_name\}} \\

\texttt{spec.trace\_id\_field} 
& Trace ID field 
& Identifies an end-to-end request trace \\

\texttt{spec.span\_id\_field} 
& Span ID field 
& Identifies an individual span \\

\texttt{spec.parent\_span\_id\_field} 
& Parent span ID field 
& Used to construct the call hierarchy \\

\texttt{spec.protocol} 
& Tracing protocol 
& \texttt{jaeger}, \texttt{zipkin}, \texttt{otlp} \\

\texttt{spec.time\_field} 
& Time field 
& Typically set to \texttt{\_\_time\_\_} \\

\texttt{spec.fields} 
& Field definitions 
& Includes all trace-related fields \\

\bottomrule
\end{tabular}
}
\end{table*}

\subsection{Entity Tools}
\label{sec:entitytools}

Details shown in Table \ref{tab:entity_tools}.


\begin{tcolorbox}[
  title={C.4: Data Link Schema},
  colback=white,
  colframe=black,
  float,
  floatplacement=p,
  label={box:entityset}
]
\begin{lstlisting}[basicstyle=\ttfamily\scriptsize]
kind: data_link
metadata:
  name: "aiops.service_related_to_aiops.metric.service"
  display_name:
    zh_cn: "service-service.metric"
  domain: aiops
spec:
  src:
    domain: aiops
    kind: entity_set
    name: aiops.service
  dest:
    domain: aiops
    kind: metric_set
    name: aiops.metric.service
  data_link_type: related_to
  fields_mapping:
    "service": "service"
\end{lstlisting}
\end{tcolorbox}

\begin{table*}[htbp]
\centering
\caption{Entity-oriented tooling interface.}
\label{tab:entity_tools}
\resizebox{\linewidth}{!}{
\begin{tabular}{lll}
\toprule
\textbf{Objective} & \textbf{Tool} & \textbf{Description} \\
\midrule

\multirow{2}{*}{Entity Information} 
& \texttt{get\_entities} 
& Retrieve a detailed list of entities \\

& \texttt{get\_neighbor\_entities} 
& Retrieve adjacent or related entities \\

\midrule

\multirow{6}{*}{Entity Data} 
& \texttt{get\_golden\_metrics} 
& Retrieve multiple key (golden) metrics of an entity \\

& \texttt{get\_metric}/\texttt{label\_values} 
& Retrieve a specific metric or label value of an entity \\

& \texttt{get\_log}/\texttt{trace}/\texttt{profile} 
& Retrieve logs, traces, or profiling data of an entity \\

& \texttt{get\_relation\_golden\_metrics} 
& Retrieve multiple key metrics of an entity relationship \\

& \texttt{get\_relation\_metric}/\texttt{label\_values} 
& Retrieve a specific metric or label value of an entity relationship \\

& \texttt{get\_relation\_log}/\texttt{trace} 
& Retrieve logs or traces of an entity relationship \\

\midrule

\multirow{3}{*}{Knowledge and Actions} 
& \texttt{search\_knowledge} 
& Retrieve knowledge related to an entity \\

& \texttt{do\_observation} 
& Execute an observation task on an entity \\

& \texttt{do\_action} 
& Execute an action on an entity \\

\bottomrule
\end{tabular}
}
\end{table*}

%% file: reference.bib
@misc{google_antigravity_2025,
  author = {Google LLC},
  title = {Google Antigravity: Next-Generation Agentic Development Platform},
  howpublished = {\url{https://antigravity.google/}},
  year = {2025},
  organization = {Google LLC}
}

@misc{replit_2024,
  author = {Replit, Inc.},
  title = {Replit: Online IDE \& Code Editor for Every Language},
  howpublished = {\url{https://replit.com/}},
  year = {2024},
  organization = {Replit, Inc.}
}

@misc{gartner_aiops_2016,
title = {Gartner Market Guide for AIOps: Essential Reading for ITOps and SRE},
author = {IBM},
year = {2026},
month = feb,
  url          = {https://www.ibm.com/think/insights/gartner-market-guide-for-aiops-essential-reading-for-itops-and-sre}
}

@misc{gartner_aiops_maturity_2025,
  author = {Gartner},
  title = {AIOps Maturity Model: From Automation to Autonomous IT Operations},
  year = {2025},
  url  = {https://www.gartner.com/en/information-technology/glossary/aiops-artificial-intelligence-operations},
}

@article{diaz_de_arcaya_2023_aiops_mlops_survey,
  title={A joint study of the challenges, opportunities, and roadmap of mlops and aiops: A systematic survey},
  author={Diaz-De-Arcaya, Josu and Torre-Bastida, Ana I and Z{\'a}rate, Gorka and Mi{\~n}{\'o}n, Ra{\'u}l and Almeida, Aitor},
  journal={ACM Computing Surveys},
  volume={56},
  number={4},
  pages={1--30},
  year={2023},
  publisher={ACM New York, NY, USA}
}

@article{zhang_2024_aiops_llm_survey,
  title={A survey of aiops for failure management in the era of large language models},
  author={Zhang, Lingzhe and Jia, Tong and Jia, Mengxi and Wu, Yifan and Liu, Aiwei and Yang, Yong and Wu, Zhonghai and Hu, Xuming and Yu, Philip S and Li, Ying},
  journal={arXiv preprint arXiv:2406.11213},
  year={2024}
}

@misc{rfc5424_syslog,
  title        = {The Syslog Protocol},
  author       = {Gerhards, R.},
  year         = {2009},
  howpublished = {RFC 5424},
  url          = {https://www.rfc-editor.org/rfc/rfc5424}
}

@misc{rfc3954_netflow9,
  title        = {Cisco Systems NetFlow Services Export Version 9},
  author       = {Claise, B.},
  year         = {2004},
  howpublished = {RFC 3954},
  url          = {https://www.rfc-editor.org/rfc/rfc3954}
}

@misc{rfc3176_sflow,
  title        = {InMon Corporation's sFlow: A Method for Monitoring Traffic in Switched and Routed Networks},
  author       = {Phaal, P. and Panchen, S. and McKee, N.},
  year         = {2001},
  howpublished = {RFC 3176},
  url          = {https://www.rfc-editor.org/rfc/rfc3176}
}

@article{p4_2014,
  title   = {P4: Programming Protocol-Independent Packet Processors},
  author  = {Bosshart, Pat and Gibb, Glen and Kim, Hunseok and Varghese, George and McKeown, Nick and Izzard, Martin and Mujica, Fernando and Horowitz, Mark},
  journal = {ACM SIGCOMM Computer Communication Review},
  volume  = {44},
  number  = {3},
  pages   = {87--95},
  year    = {2014},
  doi     = {10.1145/2656877.2656890}
}

@inproceedings{bpf1993,
  title     = {The BSD Packet Filter: A New Architecture for User-level Packet Capture},
  author    = {McCanne, Steven and Jacobson, Van},
  booktitle = {Proceedings of the USENIX Winter Conference},
  year      = {1993},
  url       = {https://www.tcpdump.org/papers/bpf-usenix93.pdf}
}

@misc{ebpf_runtime_2024,
  title        = {The eBPF Runtime in the Linux Kernel},
  author       = {Alabi, Timothy and others},
  year         = {2024},
  howpublished = {arXiv preprint},
  eprint       = {2410.00026},
  archivePrefix= {arXiv},
  primaryClass = {cs.OS},
  url          = {https://arxiv.org/abs/2410.00026}
}

@inproceedings{cui2025tshape,
  title={TShape: Rescuing Machine Learning Models from Complex Shapelet Anomalies},
  author={Cui, Hang and Li, Jingjing and Si, Haotian and Zhou, Quan and Pei, Changhua and Xie, Gaogang and Pei, Dan},
  booktitle={2025 IEEE 36th International Symposium on Software Reliability Engineering Workshops (ISSREW)},
  pages={9--14},
  year={2025},
  organization={IEEE}
}

@inproceedings{omnianomaly2019,
  title     = {Robust Anomaly Detection for Multivariate Time Series through Stochastic Recurrent Neural Network},
  author    = {Su, Ya and Zhao, Youjian and Niu, Chenjie and Liu, Rong and Sun, Wei and Pei, Dan},
  booktitle = {Proceedings of KDD},
  year      = {2019},
  doi       = {10.1145/3292500.3330672}
}

@inproceedings{anomaly_transformer2021,
  title     = {Anomaly Transformer: Time Series Anomaly Detection with Association Discrepancy},
  author    = {Xu, Jie and Wu, Haixu and Wang, Jianmin and Long, Mingsheng},
  booktitle = {International Conference on Learning Representations (ICLR)},
  year      = {2022},
  url       = {https://openreview.net/forum?id=LzQQ89U1qm_}
}

@inproceedings{drain2017,
  title     = {Drain: An Online Log Parsing Approach with Fixed Depth Tree},
  author    = {He, Pinjia and Zhu, Jieming and He, Shilin and Li, Jian and Lyu, Michael R.},
  booktitle = {Proceedings of ICWS},
  year      = {2017},
  doi       = {10.1109/ICWS.2017.13}
}

@inproceedings{deeplog2017,
  title     = {DeepLog: Anomaly Detection and Diagnosis from System Logs through Deep Learning},
  author    = {Du, Min and Li, Feifei and Zheng, Guineng and Srikumar, Vivek},
  booktitle = {Proceedings of CCS},
  year      = {2017},
  doi       = {10.1145/3133956.3134015}
}

@inproceedings{logbert2022,
  title     = {LogBERT: Log Anomaly Detection via BERT},
  author    = {Guo, Haixu and Yuan, Shuhan and Wu, Jianing and others},
  booktitle = {Proceedings of IJCNN},
  year      = {2022},
  doi       = {10.1109/IJCNN55064.2022.9892280}
}

@inproceedings{microrca2020,
  title     = {MicroRCA: Root Cause Localization of Performance Issues in Microservices},
  author    = {Wu, Li and Tordsson, Johan and Elmroth, Erik and Kao, Odej},
  booktitle = {IEEE/IFIP Network Operations and Management Symposium (NOMS)},
  year      = {2020},
  doi       = {10.1109/NOMS47738.2020.9110353},
  url       = {https://github.com/elastisys/MicroRCA}
}

@inproceedings{microcause2020,
  title     = {Localizing Failure Root Causes in a Microservice through Causality Inference},
  author    = {Meng, Yuan and Zhang, Shenglin and Sun, Yongqian and Zhang, Ruru and Hu, Zhilong and Zhang, Yiyin and Jia, Chenyang and Wang, Zhaogang and Pei, Dan},
  booktitle = {Proceedings of IEEE/ACM IWQoS},
  year      = {2020},
  url       = {https://nkcs.iops.ai/wp-content/uploads/2020/07/paper-IWQOS2020-MicroCause.pdf}
}

@inproceedings{microhecl2021,
  title     = {MicroHECL: High-Efficient Root Cause Localization in Large-Scale Microservice Systems},
  author    = {Liu, Dewei and He, Chuan and Peng, Xin and Lin, Fan and Zhang, Chenxi and Gong, Shengfang and Li, Ziang and Ou, Jiayu and Wu, Zheshun},
  booktitle = {ICSE-SEIP},
  year      = {2021},
  doi       = {10.1109/ICSE-SEIP52600.2021.00043},
  url       = {https://arxiv.org/abs/2103.01782}
}

@misc{diagfusion2023,
  title        = {Robust Failure Diagnosis of Microservice System through Multimodal Data},
  author       = {Zhang, Shenglin and Jin, Pengxiang and Lin, Zihan and Sun, Yongqian and Zhang, Bicheng and Xia, Sibo and Li, Zhengdan and Zhong, Zhenyu and Ma, Minghua and Jin, Wa and Zhang, Dai and Zhu, Zhenyu and Pei, Dan},
  year         = {2023},
  howpublished = {arXiv preprint},
  eprint       = {2302.10512},
  archivePrefix= {arXiv},
  primaryClass = {cs.SE},
  doi          = {10.48550/arXiv.2302.10512},
  url          = {https://arxiv.org/abs/2302.10512}
}

@inproceedings{nedelkoski2019,
  title     = {Anomaly Detection from System Tracing Data using Multimodal Deep Learning},
  author    = {Nedelkoski, Sasho and Cardoso, Jorge and Kao, Odej},
  booktitle = {IEEE International Conference on Cloud Computing (CLOUD)},
  year      = {2019},
  doi       = {10.1109/CLOUD.2019.00064}
}

@misc{react2023,
  title        = {ReAct: Synergizing Reasoning and Acting in Language Models},
  author       = {Yao, Shunyu and Zhao, Jeffrey and Yu, Dian and Du, Nan and Shafran, Izhak and Narasimhan, Karthik and others},
  year         = {2023},
  howpublished = {arXiv preprint},
  eprint       = {2210.03629},
  archivePrefix= {arXiv},
  url          = {https://arxiv.org/abs/2210.03629}
}

@misc{toolformer2023,
  title        = {Toolformer: Language Models Can Teach Themselves to Use Tools},
  author       = {Schick, Timo and Dwivedi-Yu, Jane and Dess{\`i}, Roberto and Raileanu, Roberta and others},
  year         = {2023},
  howpublished = {arXiv preprint},
  eprint       = {2302.04761},
  archivePrefix= {arXiv},
  url          = {https://arxiv.org/abs/2302.04761}
}

@misc{llm4aiops_survey2025,
  title        = {A Survey of AIOps in the Era of Large Language Models},
  author       = {Zhang, Lingzhe and Jia, Tong and Jia, Mengxi and Wu, Yifan and Liu, Aiwei and Yang, Yong and Wu, Zhonghai and Hu, Xuming and Yu, Philip S. and Li, Ying},
  year         = {2025},
  howpublished = {arXiv preprint},
  eprint       = {2507.12472},
  archivePrefix= {arXiv},
  primaryClass = {cs.SE},
  doi          = {10.48550/arXiv.2507.12472},
  url          = {https://arxiv.org/abs/2507.12472}
}

@inproceedings{nie2025dest,
  title={DeST: An Unsupervised Decoupled Spatio-Temporal Framework for Microservice Incident Management},
  author={Nie, Xiaohui and Cui, Hang and Pei, Changhua and Si, Haotian and Xiang, Ke and Li, Jingjing and Li, Yanbiao and Xie, Gaogang and Pei, Dan},
  booktitle={2025 IEEE 36th International Symposium on Software Reliability Engineering (ISSRE)},
  pages={335--346},
  year={2025},
  organization={IEEE}
}

@inproceedings{xu2018unsupervised,
  title={Unsupervised anomaly detection via variational auto-encoder for seasonal kpis in web applications},
  author={Xu, Haowen and Chen, Wenxiao and Zhao, Nengwen and Li, Zeyan and Bu, Jiahao and Li, Zhihan and Liu, Ying and Zhao, Youjian and Pei, Dan and Feng, Yang and others},
  booktitle={Proceedings of the 2018 world wide web conference},
  pages={187--196},
  year={2018}
}

@inproceedings{wang2024revisiting,
  title={Revisiting vae for unsupervised time series anomaly detection: A frequency perspective},
  author={Wang, Zexin and Pei, Changhua and Ma, Minghua and Wang, Xin and Li, Zhihan and Pei, Dan and Rajmohan, Saravan and Zhang, Dongmei and Lin, Qingwei and Zhang, Haiming and others},
  booktitle={Proceedings of the ACM web conference 2024},
  pages={3096--3105},
  year={2024}
}

@inproceedings{li2022actionable,
  title={Actionable and interpretable fault localization for recurring failures in online service systems},
  author={Li, Zeyan and Zhao, Nengwen and Li, Mingjie and Lu, Xianglin and Wang, Lixin and Chang, Dongdong and Nie, Xiaohui and Cao, Li and Zhang, Wenchi and Sui, Kaixin and others},
  booktitle={Proceedings of the 30th ACM Joint European Software Engineering Conference and Symposium on the Foundations of Software Engineering},
  pages={996--1008},
  year={2022}
}

@article{soldani2022anomaly,
  title={Anomaly detection and failure root cause analysis in (micro) service-based cloud applications: A survey},
  author={Soldani, Jacopo and Brogi, Antonio},
  journal={ACM Computing Surveys (CSUR)},
  volume={55},
  number={3},
  pages={1--39},
  year={2022},
  publisher={ACM New York, NY}
}

@inproceedings{yu2023nezha,
  title={Nezha: Interpretable fine-grained root causes analysis for microservices on multi-modal observability data},
  author={Yu, Guangba and Chen, Pengfei and Li, Yufeng and Chen, Hongyang and Li, Xiaoyun and Zheng, Zibin},
  booktitle={Proceedings of the 31st ACM Joint European Software Engineering Conference and Symposium on the Foundations of Software Engineering},
  pages={553--565},
  year={2023}
}

@inproceedings{wang2024rcagent,
  title={Rcagent: Cloud root cause analysis by autonomous agents with tool-augmented large language models},
  author={Wang, Zefan and Liu, Zichuan and Zhang, Yingying and Zhong, Aoxiao and Wang, Jihong and Yin, Fengbin and Fan, Lunting and Wu, Lingfei and Wen, Qingsong},
  booktitle={Proceedings of the 33rd ACM International Conference on Information and Knowledge Management},
  pages={4966--4974},
  year={2024}
}

@misc{prometheus,
  author       = {{Prometheus Team}},
  title        = {{Prometheus: Monitoring system and time series database}},
  howpublished = {\url{https://prometheus.io/}},
  year         = {2012},
  note         = {Accessed: 2025-08-21}
}

@misc{elasticsearch,
  author       = {{Elastic NV}},
  title        = {{Elasticsearch: Open Source Distributed RESTful Search and Analytics Engine}},
  howpublished = {\url{https://www.elastic.co/elasticsearch}},
  year         = {2010},
  note         = {Accessed: 2025-08-21}
}

@inproceedings{sun2025aiopsarena,
  title={AIOpsArena: Scenario-Oriented Evaluation and Leaderboard for AIOps Algorithms in Microservices},
  author={Sun, Yongqian and Wang, Jiaju and Li, Zhengdan and Nie, Xiaohui and Ma, Minghua and Zhang, Shenglin and Ji, Yuhe and Zhang, Lu and Long, Wen and Chen, Hengmao and others},
  booktitle={2025 IEEE International Conference on Software Analysis, Evolution and Reengineering (SANER)},
  pages={809--813},
  year={2025},
  organization={IEEE}
}

@inproceedings{zhou2025kanad,
  title={KAN-AD: Time Series Anomaly Detection with Kolmogorov-Arnold Networks},
  author={Zhou, Quan and Pei, Changhua and Sun, Fei and Jing, Han and Gao, Zhengwei and Zhang, Haiming and Xie, Gaogang and Pei, Dan and Li, Jianhui},
  booktitle={Proceedings of the 42nd International Conference on Machine Learning (ICML)},
  year={2025},
  url={https://openreview.net/forum?id=LWQ4zu9SdQ}
}

@inproceedings{xie2023point,
  title={From point-wise to group-wise: A fast and accurate microservice trace anomaly detection approach},
  author={Xie, Zhe and Pei, Changhua and Li, Wanxue and Jiang, Huai and Su, Liangfei and Li, Jianhui and Xie, Gaogang and Pei, Dan},
  booktitle={Proceedings of the 31st ACM Joint European Software Engineering Conference and Symposium on the Foundations of Software Engineering},
  pages={1739--1749},
  year={2023}
}

@article{zhang2025failure,
  title={Failure diagnosis in microservice systems: A comprehensive survey and analysis},
  author={Zhang, Shenglin and Xia, Sibo and Fan, Wenzhao and Shi, Binpeng and Xiong, Xiao and Zhong, Zhenyu and Ma, Minghua and Sun, Yongqian and Pei, Dan},
  journal={ACM Transactions on Software Engineering and Methodology},
  volume={35},
  number={1},
  pages={1--55},
  year={2025},
  publisher={ACM New York, NY}
}

@article{yu2026survey,
  title={A survey on failure analysis and fault injection in AI systems},
  author={Yu, Guangba and Tan, Gou and Huang, Haojia and Zhang, Zhenyu and Chen, Pengfei and Natella, Roberto and Zheng, Zibin and Lyu, Michael R},
  journal={ACM Transactions on Software Engineering and Methodology},
  volume={35},
  number={1},
  pages={1--42},
  year={2026},
  publisher={ACM New York, NY}
}

@inproceedings{pei2025flow,
  title={Flow-of-Action: SOP Enhanced LLM-Based Multi-Agent System for Root Cause Analysis},
  author={Pei, Changhua and Wang, Zexin and Liu, Fengrui and Li, Zeyan and Liu, Yang and He, Xiao and Kang, Rong and Zhang, Tieying and Chen, Jianjun and Li, Jianhui and others},
  booktitle={Companion Proceedings of the ACM on Web Conference 2025},
  pages={422--431},
  year={2025}
}

@article{hou2025model,
  title={Model context protocol (mcp): Landscape, security threats, and future research directions},
  author={Hou, Xinyi and Zhao, Yanjie and Wang, Shenao and Wang, Haoyu},
  journal={arXiv preprint arXiv:2503.23278},
  year={2025}
}

@article{kernighan1979unix,
  title={The UNIX™ programming environment},
  author={Kernighan, Brian W and Mashey, John R},
  journal={Software: Practice and Experience},
  volume={9},
  number={1},
  pages={1--15},
  year={1979},
  publisher={Wiley Online Library}
}

@article{zhang2025promassistant,
  title={PromAssistant: Leveraging Large Language Models for Text-to-PromQL},
  author={Zhang, Chenxi and Zhang, Bicheng and Yang, Dingyu and Peng, Xin and Chen, Miao and Xie, Senyu and Chen, Gang and Bi, Wei and Li, Wei},
  journal={arXiv preprint arXiv:2503.03114},
  year={2025}
}

@inproceedings{ding2024rd,
  title={Rd-probe: Scalable monitoring with sufficient coverage in complex datacenter networks},
  author={Ding, Rui and Liu, Xunpeng and Yang, Shibo and Huang, Qun and Xie, Baoshu and Sun, Ronghua and Zhang, Zhi and Cui, Bolong},
  booktitle={Proceedings of the ACM SIGCOMM 2024 Conference},
  pages={258--273},
  year={2024}
}

@inproceedings{zheng2024mumon,
  title={$\mu$Mon: Empowering Microsecond-level Network Monitoring with Wavelets},
  author={Zheng, Hao and Huang, Chengyuan and Han, Xiangyu and Zheng, Jiaqi and Wang, Xiaoliang and Tian, Chen and Dou, Wanchun and Chen, Guihai},
  booktitle={Proceedings of the ACM SIGCOMM 2024 Conference},
  pages={274--290},
  year={2024}
}

@inproceedings{chen2024eagle,
  title={Eagle: Toward scalable and near-optimal network-wide sketch deployment in network measurement},
  author={Chen, Xiang and Xiao, Qingjiang and Liu, Hongyan and Huang, Qun and Zhang, Dong and Liu, Xuan and Hu, Longbing and Zhou, Haifeng and Wu, Chunming and Ren, Kui},
  booktitle={Proceedings of the ACM SIGCOMM 2024 Conference},
  pages={291--310},
  year={2024}
}

@inproceedings{mehner2024ipd,
  title={IPD: Detecting Traffic Ingress Points at ISPs},
  author={Mehner, Stefan and Reelfs, Helge and Poese, Ingmar and Hohlfeld, Oliver},
  booktitle={Proceedings of the ACM SIGCOMM 2024 Conference},
  pages={778--793},
  year={2024}
}

@inproceedings{zhang2025achieving,
  title={Achieving High-Speed and Robust Encrypted Traffic Anomaly Detection with Programmable Switches},
  author={Zhang, Han and Liu, Guyue and Shi, Xingang and Li, Yahui and He, Dongbiao and Wang, Jilong and Wang, Zhiliang and Zhu, Yongqing and Ruan, Ke and Cao, Weihua and others},
  booktitle={Proceedings of the ACM SIGCOMM 2025 Conference},
  pages={1254--1256},
  year={2025}
}

@inproceedings{wang2025hawkeye,
  title={Hawkeye: Diagnosing rdma network performance anomalies with pfc provenance},
  author={Wang, Shicheng and Zhang, Menghao and Li, Xiao and Peng, Qiyang and Yu, Haoyuan and Wang, Zhiliang and Xu, Mingwei and Hu, Xiaohe and Yang, Jiahai and Shi, Xingang},
  booktitle={Proceedings of the ACM SIGCOMM 2025 Conference},
  pages={481--495},
  year={2025}
}

@inproceedings{sundara2023global,
  title={Global, passive detection of connection tampering},
  author={Sundara Raman, Ram and Merino, Louis-Henri and Bock, Kevin and Fayed, Marwan and Levin, Dave and Sullivan, Nick and Valenta, Luke},
  booktitle={Proceedings of the ACM SIGCOMM 2023 Conference},
  pages={622--636},
  year={2023}
}

@inproceedings{wang2025towards,
  title={Towards llm-based failure localization in production-scale networks},
  author={Wang, Chenxu and Zhang, Xumiao and Lu, Runwei and Lin, Xianshang and Zeng, Xuan and Zhang, Xinlei and An, Zhe and Wu, Gongwei and Gao, Jiaqi and Tian, Chen and others},
  booktitle={Proceedings of the ACM SIGCOMM 2025 Conference},
  pages={496--511},
  year={2025}
}

@inproceedings{liu2025skeletonhunter,
  title={SkeletonHunter: Diagnosing and Localizing Network Failures in Containerized Large Model Training},
  author={Liu, Wei and Qian, Kun and Li, Zhenhua and Xu, Tianyin and Liu, Yunhao and Wang, Weicheng and Zhang, Yun and Li, Jiakang and Zhu, Shuhong and Li, Xue and others},
  booktitle={Proceedings of the ACM SIGCOMM 2025 Conference},
  pages={527--540},
  year={2025}
}

@inproceedings{yang2025skynet,
  title={SkyNet: Analyzing Alert Flooding from Severe Network Failures in Large Cloud Infrastructures},
  author={Yang, Bo and Hu, Huanwu and Li, Yifan and Li, Yunguang and Tang, Xiangyu and Tian, Bingchuan and Wu, Gongwei and Xu, Jianfeng and Zhang, Xumiao and Chen, Feng and others},
  booktitle={Proceedings of the ACM SIGCOMM 2025 Conference},
  pages={512--526},
  year={2025}
}

@inproceedings{francis2018cypher,
  title={Cypher: An evolving query language for property graphs},
  author={Francis, Nadime and Green, Alastair and Guagliardo, Paolo and Libkin, Leonid and Lindaaker, Tobias and Marsault, Victor and Plantikow, Stefan and Rydberg, Mats and Selmer, Petra and Taylor, Andr{\'e}s},
  booktitle={Proceedings of the 2018 international conference on management of data},
  pages={1433--1445},
  year={2018}
}

@article{shinn2023reflexion,
  title={Reflexion: Language agents with verbal reinforcement learning},
  author={Shinn, Noah and Cassano, Federico and Gopinath, Ashwin and Narasimhan, Karthik and Yao, Shunyu},
  journal={Advances in neural information processing systems},
  volume={36},
  pages={8634--8652},
  year={2023}
}

@inproceedings{wang2023plan,
  title={Plan-and-solve prompting: Improving zero-shot chain-of-thought reasoning by large language models},
  author={Wang, Lei and Xu, Wanyu and Lan, Yihuai and Hu, Zhiqiang and Lan, Yunshi and Lee, Roy Ka-Wei and Lim, Ee-Peng},
  booktitle={Proceedings of the 61st annual meeting of the association for computational linguistics (volume 1: long papers)},
  pages={2609--2634},
  year={2023}
}

@inproceedings{chan2024chateval,
  title={Chateval: Towards better llm-based evaluators through multi-agent debate},
  author={Chan, Chi-Min and Chen, Weize and Su, Yusheng and Yu, Jianxuan and Xue, Wei and Zhang, Shanghang and Fu, Jie and Liu, Zhiyuan},
  booktitle={International conference on learning representations},
  volume={2024},
  pages={9079--9093},
  year={2024}
}

@misc{openclaw2026,
  title        = {{OpenClaw: Personal AI Assistant}},

  howpublished = {\url{https://github.com/openclaw/openclaw}},

}
